%% file: paper.tex
\documentclass[letterpaper,twocolumn,10pt,dvipsnames]{article}
\usepackage{usenix2019_v3}

\usepackage{amsmath,amsfonts}
\usepackage[utf8]{inputenx}
\usepackage[T1]{fontenc}
\usepackage{marvosym}
\usepackage{graphicx}
\usepackage{textcomp}
\usepackage{url}
\usepackage{subcaption}
\usepackage{paralist}
\usepackage{wrapfig}
\usepackage{multirow}
\usepackage{listings}
\usepackage{booktabs}
\usepackage{makecell}
\usepackage{boxedminipage}
\usepackage{amsmath}
\usepackage{graphicx}
\usepackage{minibox}
\usepackage{algorithm2e}
\usepackage{algpseudocode}
\usepackage{pifont}
\usepackage{caption}
\usepackage{fancyhdr}
\usepackage{lipsum}
\let\labelindent\relax
\usepackage{enumitem}
\usepackage{subfiles}
\usepackage[english]{babel}
\usepackage[]{ifsym}
\usepackage{amsthm}

\usepackage{tikz}
\usepackage{stmaryrd}
\usepackage{mathtools}
\usepackage{cleveref}
\usepackage[framemethod=TikZ]{mdframed} % box
\usepackage{MnSymbol}
\usepackage{tabularx}  % for table
\usepackage{adjustbox} 
\usepackage{array}   
\usepackage{changes}
\usepackage[all]{nowidow}
\usepackage{upgreek}
\usepackage{wasysym}
\usepackage{pifont}
\usepackage{authblk}
\newcommand{\xmarkn}[0]{\ding{53}}

\usepackage{ulem}
\usepackage{hyperref}
\AtBeginDocument{%
  }

\newcommand{\blfootnote}[1]{%
  \begingroup
  \renewcommand{\thefootnote}{}% Remove footnote number
  \footnote{#1}%
  \addtocounter{footnote}{-1}% Reset footnote counter
  \endgroup
}
\begin{document}
\input{macro}

% \title{\sysname: Safe CVM-Hypervisor Coordination for Dynamic CPU Resources Management}

% \title{Boosting Microservices: Secure and On-demand CPU Allocation for Confidential VMs}% with Untrustworthy Hypervisors}

% \title{Elastic CVM and \workcpu: Secure and On-demand Runtime CPU Adjustment for CVM}

\title{Elastic Confidential VM: Enabling Secure and On-demand Runtime CPU Adjustment}

\title{Elastic Confidential VM: Secure and On-Demand CPU Scaling}

\title{On-Demand CPU Adjustment in Secure VMs: Enabling Elastic Confidential Computing}

\title{Elastic Confidential VMs with Secure and On-Demand CPU Scaling}

\title{\sysname: Elastic Confidential VMs with Secure and Dynamic CPU Scaling}

% we define worker vcpu to enable a elastic CVM model. Since the second part of the title mentions it is a CPU adjustment, maybe highlight elastic CVM model and the worker vCPU in the first part is better?
% \title{Elastic vCPU: Secure and On-demand Runtime CPU Adjustment for Confidential VMs}

\author{
    {\rm Shixuan Zhao*\textsuperscript{1}} 
    {\rm Mengyuan Li*\textsuperscript{2,3}} 
    {\rm Mengjia Yan\textsuperscript{3}}
    {\rm Zhiqiang Lin\textsuperscript{1}}
}

\affil{\textsuperscript{1}The Ohio State University, 
\textsuperscript{2}University of Southern California, 
\textsuperscript{3}Massachusetts Institute of Technology}

\maketitle
\blfootnote{*The two authors contributed equally to this paper.}
\begin{abstract}
% All figure sources can be found here:  google drive and draw.io
% https://drive.google.com/file/d/1_4UQCmny0ACkZba48TYbPf32QBgi0ts_/view?usp=sharing

Confidential Virtual Machines (CVMs) are a type of VM-based Trusted Execution Environments (TEEs) designed to enhance the security of cloud-based VMs, safeguarding them even from malicious hypervisors. Although CVMs have been widely adopted by major cloud service providers, current CVM designs face significant challenges in runtime resource management due to their fixed capacities and lack of transparency. These limitations hamper efficient cloud resource management, leading to increased operational costs and reduced agility in responding to fluctuating workloads.

This paper introduces a dynamic CPU resource management approach, featuring the novel concept of ``Elastic CVM''. This approach allows for hypervisor-assisted runtime adjustment of CPU resources using a specialized vCPU type, termed \workcpu. This new approach enhances CPU resource adaptability and operational efficiency without compromising security. Additionally, we introduce a \workcpu Abstraction Layer to simplify \workcpu deployment and management. 
To demonstrate the effectiveness of our approach, we have designed and implemented a serverless computing prototype platform, called \sysname. We show that \sysname significantly improves performance and efficiency through finer-grain resource management. The concept of ``Elastic CVM'' and the \workcpu design not only optimize cloud resource utilization but also pave the way for more flexible and cost-effective confidential computing environments.

\end{abstract}

\input{1-intro}
\input{2-background}

\input{3-overview}

\input{4-details}

\input{5-evaluation}
\input{6-discussion}

\input{7-related}

\input{8-conclusion}

\input{paper.bbl}

\input{0-appendix}

\end{document}

%% file: macro.tex
\newcommand{\bheading}[1]{{\vspace{4pt}\noindent{\textbf{#1}}}}
% \newcolumntype{?}{!{\vrule width 1pt}}

\newcommand{\xmark}{\ding{55}}%

\newcommand{\tikzcircle}[2][red,fill=red]{\tikz[baseline=-0.5ex]\draw[#1,radius=#2] (0,0) circle ;}%

\newcommand*\emptycirc[1][0.7ex]{\tikz\draw (0,0) circle (#1);} 
\newcommand*\halfcirc[1][0.7ex]{%
  \begin{tikzpicture}
  \draw[fill] (0,0)-- (90:#1) arc (90:270:#1) -- cycle ;
  \draw (0,0) circle (#1);
  \end{tikzpicture}}
\newcommand*\fullcirc[1][0.7ex]{\tikz\fill (0,0) circle (#1);}

\newcommand{\draft}[1]{\textcolor{blue}{#1}}
%terminology and formatting
\newcounter{note}[section]
\renewcommand{\thenote}{\thesection.\arabic{note}}

\newcommand{\mengyuan}[1]{\refstepcounter{note}{\bf\textcolor{olive}{$\ll$mengyuan~\thenote: {\sf #1}$\gg$}}}

\newcommand{\figurewidth}{\columnwidth}
\newcommand{\secref}[1]{\mbox{Section~\ref{#1}}\xspace}
\newcommand{\secrefs}[2]{\mbox{Section~\ref{#1}--\ref{#2}}\xspace}
\newcommand{\figref}[1]{\mbox{Figure~\ref{#1}}}
\newcommand{\algrref}[1]{\mbox{Algorithm~\ref{#1}}}
\newcommand{\tabref}[1]{\mbox{Table~\ref{#1}}}
\newcommand{\appref}[1]{\mbox{Appendix~\ref{#1}}}
\newcommand{\ignore}[1]{}

% writing styles
\newcommand{\etc}{\textit{etc.}\xspace}
\newcommand{\ie}{\textit{i.e.}\xspace}
\newcommand{\eg}{\textit{e.g.}\xspace}
\newcommand{\aka}{\textit{a.k.a.}\xspace}
\newcommand{\etal}{\textit{et al.}\xspace}
\newcommand{\tabincell}[2]{\begin{tabular}{@{}#1@{}}#2\end{tabular}}
% terms

%\newcommand{\ignore}[1]{}

\newcommand{\sysname}{\textsc{Ditto}\xspace}
\newcommand{\workcpu}{Worker vCPU\xspace}
\newcommand{\workcpus}{Worker vCPUs\xspace}
\newcommand{\workcpunospace}{Worker vCPU}
\newcommand{\workcpusnospace}{Worker vCPUs}

\renewcommand{\todo}[1]{\refstepcounter{note}{\bf\textcolor{red}{$\ll$ToFill~\thenote: {\sf #1}$\gg$}}}

\newcommand{\addr}[1]{\ensuremath{P_{#1}}\xspace}
\newcommand{\offset}[1]{\ensuremath{[O_{#1}]}\xspace}
\newcommand{\ts}[1]{\ensuremath{S_{#1}}\xspace}
\newcommand{\gCR}[1]{gCR3\ensuremath{_{#1}}\xspace}
% units
\newcommand{\gbytes}{\ensuremath{\mathrm{GB}}\xspace}
\newcommand{\mbytes}{\ensuremath{\mathrm{MB}}\xspace}
\newcommand{\kbytes}{\ensuremath{\mathrm{KB}}\xspace}
\newcommand{\bytes}{\ensuremath{\mathrm{B}}\xspace}
\newcommand{\hertz}{\ensuremath{\mathrm{Hz}}\xspace}
\newcommand{\ghertz}{\ensuremath{\mathrm{GHz}}\xspace}
\newcommand{\msecs}{\ensuremath{\mathrm{ms}}\xspace}
\newcommand{\usecs}{\ensuremath{\mathrm{\mu{}s}}\xspace}
\newcommand{\nsecs}{\ensuremath{\mathrm{ns}}\xspace}
\newcommand{\secs}{\ensuremath{\mathrm{s}}\xspace}
\newcommand{\gbits}{\ensuremath{\mathrm{Gb}}\xspace}

\newcounter{packednmbr}
\newenvironment{packedenumerate}{
\begin{list}{\thepackednmbr.}{\usecounter{packednmbr}
\setlength{\itemsep}{0pt}
\addtolength{\labelwidth}{4pt}
\setlength{\leftmargin}{12pt}
\setlength{\listparindent}{\parindent}
\setlength{\parsep}{3pt}
\setlength{\topsep}{3pt}}}{\end{list}}

\newenvironment{packeditemize}{
\begin{list}{$\bullet$}{
\setlength{\labelwidth}{0pt}
\setlength{\itemsep}{2pt}
\setlength{\leftmargin}{\labelwidth}
\addtolength{\leftmargin}{\labelsep}
\setlength{\parindent}{0pt}
\setlength{\listparindent}{\parindent}
\setlength{\parsep}{1pt}
\setlength{\topsep}{1pt}}}{\end{list}}

\newenvironment{methoditemize}{
\begin{list}{$\bullet$m}{
\setlength{\labelwidth}{0pt}
\setlength{\itemsep}{2pt}
\setlength{\leftmargin}{\labelwidth}
\addtolength{\leftmargin}{\labelsep}
\setlength{\parindent}{0pt}
\setlength{\listparindent}{\parindent}
\setlength{\parsep}{1pt}
\setlength{\topsep}{1pt}}}{\end{list}}

%%%% Syntax Highlight BEGIN %%%%
\lstdefinestyle{mystyle}{
  basicstyle=\ttfamily\small,
  keywordstyle={[1]\color{magenta}},
  keywordstyle={[2]\color{orange}},
  keywordstyle={[3]\color{green}},
  commentstyle=\color{OliveGreen},
  numbers=left,
  numberstyle=\tiny\color{gray},
  numbersep=3pt,
}
\lstdefinelanguage{mylang} {
    morekeywords={[1]module, method},
    morekeywords={[2]Bool, HistoryT, Addr, Data},
    morekeywords={[3]if},
    sensitive=false,
    morecomment=[l]{//},
    %morestring=[b]",
}
\lstset{style=mystyle}

% Example1
% NOTE: minipage it to avoid the code being splitted into different columns
% \begin{minipage}{\columnwidth}
% \begin{lstlisting}[style=mystyle, language=Verilog]
% module Forward(input datain, output dataout);
%   dataout = datain;
% endmodule
% \end{lstlisting}
% \end{minipage}

% Example2
% \lstinline[columns=fixed]{module Forward(input datain);}
% \lstinline[style=mystyle, language=Verilog]{module Forward(input datain);}
%%%% Syntax Highlight END %%%%

% \setlength{\abovedisplayskip}{0pt}
% \setlength{\belowdisplayskip}{0pt}
% \setlength{\abovedisplayshortskip}{0pt}
% \setlength{\belowdisplayshortskip}{0pt}

\newcommand{\hlc}[2][yellow]{{%
    \colorlet{foo}{#1}%
    \sethlcolor{foo}\hl{#2}}%
}

\mdfsetup{
  backgroundcolor=black!10,
  linecolor=black,
  roundcorner=2mm
}

%% file: 1-intro.tex
\section{Introduction}
\label{sec:intro}

A Confidential Virtual Machine (CVM) is a VM-based Trusted Execution Environment (TEE) designed to protect the security of cloud VMs. Benefiting from TEE-enforced security, CVM can safeguard the confidentiality and integrity of VM tasks even against a malicious hypervisor. Compared to process-based TEE, such as Intel Software Guard Extensions (SGX)~\cite{costan:2016:intel, vsgx}, 
CVMs have the advantages of better performance and eliminating the need for code modification. These advantages led to a surge in the development of CVM technologies, such as AMD Secure Encrypted Virtualization (SEV)~\cite{kaplan:2016:sevWpaper}, Intel Trust Domain Extensions (TDX)~\cite{intel:2020:tdx}, and ARM Confidential Compute Architecture (CCA)~\cite{arm:2021:cca}. 
Meanwhile, the main cloud service providers (CSPs), including AWS~\cite{aws:2023:sev}, Google Cloud~\cite{google:2020:sev}, and Microsoft Azure~\cite{microsoft:2021:sev}, have already incorporated CVMs into public cloud and offered commercial CVM-based  services, also known as confidential computing services.

\subsection{Motivation} 
% \sx{How about title as `Motivation'? And move the sentence as an underlined one in the main matter?}
\label{sec:intro:motivation}

\textit{There's no secure and fast runtime resource management for CVMs.} Even though the design of CVMs gives the hypervisor the ability to manage physical resources~\cite{li:2024:sok}, such as memory allocation, CPU scheduling, and I/O routing, the hypervisor cannot manage resources and CVMs as freely as it can with traditional unprotected VMs. This limitation is mainly because, for security reasons, the capacity of a CVM is usually set and fixed during initialization, and its operation during runtime is intentionally kept opaque as a \textbf{``blackbox''}. The following outline the limitations imposed by CVMs, categorized by different physical resources and specific events.
\begin{packeditemize}
\item\bheading{Slow remote attestation.} To initiate a CVM, a remote attestation is required for the TEE owner to verify the initial status before any secrets are uploaded to the CVM. However, this procedure is often costly and time-consuming, making it cumbersome to launch new CVMs for fast scaling purposes. It involves complex interactions between the CVM owner, the hypervisor, and the CPU vendor's attestation services. For instance, to protect the disk storage of a CVM, the owner may even need to upload an encrypted disk image and then provision the disk encryption key by remote attestation~\cite{amd:2019:sevapi}.
\item\bheading{Protected virtual CPU (vCPU) states.} During runtime, the status and number of vCPUs are strictly safeguarded by CPU hardware, which also adds latency and complexity to vCPU management. For instance, the vCPU register values during a \texttt{VMEXIT} are always encrypted, necessitating additional routing and protocols for instruction emulation and interrupt injection~\cite{amd:2020:ghcb}. Furthermore, the number of vCPUs assigned to a CVM is typically determined during the initialization phase for attestation purposes~\cite{tdx-no-hotplug,sev-no-hotplug}, and remains unchanged throughout runtime.
\item\bheading{Slow live migration.} Live migration is a common method used by CSP for more efficient VM resource management. However, it becomes significantly more costly with CVMs due to the necessary involvement of CPU hardware for validation and checking. For example, in AMD SEV, the secure co-processor must establish a secure transmission channel and manage CVM's memory decryption and re-encryption, creating a major bottleneck for CVM live migration~\cite{amd:2019:sevapi}.
\end{packeditemize}

Such limitations of CVMs present challenges in meeting the agile resource management needs in modern cloud computing, especially in scenarios with varying cloud workload intensities. This creates a lose-lose situation for both CSPs and CVM owners.
For CVM owners, ensuring reliable service and managing sudden demand spikes requires renting CVMs with larger capacities initially. This results in unnecessary idle resources during low-demand periods, leading to higher rental prices. 
For CSPs, managing resources becomes particularly challenging in cloud services that utilizing CVM technology. For example, in a serverless platform, CSPs must be able to efficiently adjust the size and number of physical resources assigned to each function according to workload intensity. However, confidential containers (CoCo)~\cite{coco:2024:github} nowadays are usually protected inside CVMs with fixed size. The rigidity of CVM size and slow launching time lead to the necessary of CoCo and CVM pre-allocation, increasing operational costs.
Such disadvantage of existing CVM designs raises the research question this paper aims to address: 
\begin{center}
\minibox[rule=1pt,pad=1pt]{
\begin{minipage}[h]{0.9\columnwidth}
{\it \textit{How can we dynamically and securely manage resources allocated to CVMs to adaptively meet varying workload intensities in a fast and efficient manner?}} 
\end{minipage}
}
% \vspace{1pt}
\end{center}

\subsection{\workcpu Abstraction} 
% : CVM-Hypervisor Coordination for Fast and Secure CPU Resource Management
% \sx{How about title as `\workcpu Abstraction'? And move the sentence as an underlined one in the main matter?}
\label{sec:intro:worker}

In this paper, we primarily explore the dynamic adjustment of CPU resources within CVMs.  \textit{We propose a CVM-Hypervisor coordination mechanism for fast and secure CPU resource management.}
Although the concept of dynamic vCPU adjustment~\cite{uhlig:2004:towards,mwaikambo:2004:linux,gleixner:2012:cleaning} is not new in traditional virtualization, its implementation within a CVM environment presents significant challenges. From both hardware support and software development perspectives, no existing CVM platforms currently support vCPU hotplugging, as detailed in~\secref{sec:background:hotplug}. 
Additionally, from an efficiency standpoint, dynamically managing vCPUs poses challenges for the CVM owner, who must handle complex, resource-intensive scheduling decisions with minimal support from the hypervisor, as the CVM functions as a black box.
% and could disrupt isolation within CVMs, potentially leading to abnormal behavior or even VM crashes.

% , including Intel TDX, AMD SEV, and SEV-ES, are not designed to support runtime vCPU hotplug~\cite{tdx-no-hotplug,sev-no-hotplug}. On the development side, platforms like AMD SEV-SNP and ARM CCA may have the potential to support vCPU hotplugging, but the necessary software support is either not planned or not yet available~\cite{cca-no-hotplug,amd:2020:ghcb}.

\bheading{Adaptive \workcpunospace.} To address this inefficiency and limitation, we propose the concept of ``\textit{Elastic CVM}'' and introduce a novel type of vCPU called the ``\workcpunospace'',  which is designed specifically for runtime dynamic CPU resource adjustment within CVMs. Unlike traditional vCPUs that are always active, \workcpus remain dormant when the workload of the CVM is low, conserving physical CPU resources and occupying only minimal memory to maintain their states. Moreover, each worker vCPU can be bound with specific applications. When high workloads or special events occur that regular vCPUs cannot manage, the \workcpus can be swiftly activated to directly handle the increased demand. While active, \workcpus function identically to regular vCPUs, allowing the total active vCPU count to increase and effectively accommodate workload spikes.

\bheading{\workcpu Abstraction Layer.} To ensure the security and efficiency of \workcpus, we introduce an \workcpu abstraction layer. This layer simplifies the use and deployment of \workcpu by providing clear principles for selecting suitable applications that align with the \workcpu design, as well as necessary software components for securely and efficiently deploying and managing \workcpusnospace.

For application selection, short-lived computational applications, such as compression tasks that re-recognize vCPU resources, and long-lived applications with independent worker threads, such as Nginx, serverless computing, or machine learning inferencing, can fully utilize the flexibility and rapid response offered by \workcpu designs. These types of applications also minimize the need for significant changes to the guest OS or applications.
For \textit{\workcpu management}, there are three key components to be considered.
The \textit{Deployment Module} within the CVM pre-deploys a certain number of \workcpus during initialization and assigns tasks to these \workcpusnospace. The \textit{Control Module} provides runtime control and coordinates with the hypervisor to wake up or put the \workcpus to sleep.
The \textit{Scheduling Module} determines whether a specific \workcpu should be activated or deactivated.
In this paper, we designed and implemented a hypervisor-assisted approach that delegates scheduling and dynamic adjustment of vCPU numbers to the hypervisor, offloading complex scheduling tasks and aligning perfectly with scenarios such as serverless computing. The design also ensures that the security of the CVM is not compromised.
\ignore{Depending on the applications the CVM is running, the scheduling module can reside either within the CVM or on the untrusted hypervisor.
Placing the scheduling module inside the CVM allows complete control over \workcpusnospace. In contrast, a hypervisor-controlled scheduling module delegates the dynamic adjustment of vCPU numbers to the hypervisor, offloading complex scheduling tasks and aligning perfectly with scenarios such as serverless computing.
Both configurations are designed to ensure that the security of the CVM is not compromised.}

\subsection{\sysname: Fast and Auto-regulated Serverless Prototype} 
\label{sec:intro:sys}
% \sx{Can we make this in one line?}
To illustrate the practical application of the \workcpu abstraction, we designed and implemented a confidential serverless platform prototype called \sysname. \sysname features an event-driven, auto-scaling serverless architecture that integrates Apache OpenWhisk~\cite{apache:2024:openwhisk} (an open-source serverless platform), Kubernetes~\cite{kubernetes:2024:website} (a container orchestration tool), and Confidential Containers~\cite{coco:2024:github} (based on Kata containers~\cite{kata:2024:website}).  In this typical serverless architecture, OpenWhisk operates on the hypervisor side, handling request routing and function scheduling. Kubernetes manages confidential Kata containers, while container security is protected by CVM.
Unlike existing configurations that use a fixed size per CVM, \sysname leverages the \workcpu design to enable fast and auto-regulated CVM CPU resource management. 
This design enables rapid and dynamic adjustments of CPU and worker threads based on varying workload intensities, facilitating finer-grained and quicker auto-scaling.
Consequently, it eliminates the need to rely solely on the slower process of launching new CVMs and confidential containers for new function threads.
Our evaluation in \secref{sec:evaluation} shows that our design can significantly benefit both performance and CPU efficiency when compared with a vanilla OpenWhisk system.

The contributions of our work are summarized as:
\begin{packeditemize}
\item \textbf{Elastic CVM Concept:} We introduce the concept of ``Elastic CVM'', which utilizes CVM-hypervisor coordination to dynamically adjust CPU resource allocation, enhancing the adaptability and efficiency of CVMs.

\item \textbf{Innovative \workcpu Design:} We propose a novel type of vCPU, called \workcpunospace, designed specifically for elastic operations in CVMs which can be scheduled using a hypervisor-assisted approach. \workcpus can remain dormant to conserve resources and activate quickly in response to workload increases, ensuring efficient resource management without sacrificing security.

\item \textbf{\workcpu Abstraction Layer:} We introduce the \workcpu Abstraction Layer, an abstraction layer that simplifies the deployment, management, and operation of \workcpus. It provides essential guidelines for effective \workcpu integration within CVMs. 

% \ZQ{Can the 2nd and 3rd contribution be combined, since they both are related to the innovative \workcpu design? Or you wanna callout the 3rd contribution differently?} \mengyuan{Yes, we want to highlight this}

\item \textbf{\sysname Prototype Development and Empirical Evaluation:} We developed \sysname, a confidential serverless platform  prototype that applies \workcpu design to achieve fast and auto-scaling in a serverless environment. Our empirical evaluations of \sysname indicate substantial improvements in resource utilization and operational efficiency compared to existing confidential serverless settings. 
% \ZQ{maybe highlight the "empirical evaluation" aspect in the heading?}
\end{packeditemize}

% \todo{@shixuan, after reading the paper once, please feel free to add more details in section 4. e.g., the shutdown procedure and some details about worker mask}

%% file: 2-background.tex
\section{Background}
\label{sec:background}

\subsection{Confidential Virtual Machine}
Confidential VMs (CVMs) are TEE-protected VMs that safeguard confidentiality and integrity against privileged and physical attackers. TEE enforces isolation between the CVM and the hypervisor, ensuring exclusive use of memory pages~\cite{amd:2020:snp}, encrypted memory~\cite{kaplan:2016:sevWpaper}, protected CPU registers~\cite{kaplan:2017:seves}, and some other protection mechanisms, such as restricted interrupt delivery~\cite{amd:2020:snp} and \texttt{CPUID} checks. 
CVMs can run entire operating systems within a secure boundary, offering better performance than enclave-based TEEs such as Intel SGX~\cite{costan:2016:intel}. They also support unmodified code, allowing existing applications to run directly, making them ideal for confidential computing in public clouds. 

% Notable CVM designs include AMD Secure Encrypted Virtualization (SEV)~\cite{kaplan:2016:sevWpaper}, Intel Trust Domain Extensions (TDX)~\cite{intel:2020:tdx}, and ARM Confidential Compute Architecture (CCA)~\cite{arm:2021:cca}.

% The advantage of CVMs lies in their ability to run an entire operating system within a single secure boundary. This results in better performance compared to earlier enclave-based TEE solutions, such as Intel SGX~\cite{costan:2016:intel}. Additionally, CVMs doesn't need code modification and can compile and run existing applications directly, making them a desirable choice for confidential computing in public clouds. 
% Common Commercial CVM designs include 

% Confidential VMs (CVMs) refers to hardware TEE protected VMs whose confidentiality and integrity are protected against privileged and physical attackers. 
% TEE protects CVM's data and instructions by enforcing enough isolation between the CVM and the privileged hypervisor. The isolation ensures exclusive use of memory pages~\cite{amd:2020:snp}, different keys to encrypt the memory~\cite{kaplan:2016:sevWpaper}, protected CPU register values~\cite{kaplan:2017:seves}, and some other protection mechanisms, such as restricted interrupt delivery~\cite{amd:2020:snp} and \texttt{CPUID} checks. 

\bheading{AMD Secure Encrypted Virtualization (SEV).} AMD SEV is the first commercial CVM design since 2016~\cite{kaplan:2016:sevWpaper}. Until now, there are three different generations of SEV series, with the later versions having more security checks and enhanced security features compared to the prior one. The first generation SEV does not protect CPU register values during the context switch between the CVM world and the untrusted hypervisor world, leading to a breach of security by manipulating the register backups. The second generation of SEV, so called SEV encrypted state (SEV-ES), mitigates this problem by encrypting the register backup during context switch. The latest version of SEV, so called SEV secure nested paging (SEV-SNP) introduces a reverse map table data structure (RMP table) to maintain the correct mapping information of a CVM and provide software-level memory integrity protection. 
% SEV-SNP also introduces some additional protection schemes, such as restricted interrupt injection and protected \texttt{CPUID}, which addresses most known attacks against SEV.
In this paper, we will only use an implementation prototype in SEV systems to demonstrate the effectiveness of our designs, while we believe the idea of elastic CVM and \workcpu can be widely adopted to any CVM design.

\subsection{vCPU Hotplug in CVMs}
\label{sec:background:hotplug}
vCPU hotplug is a mechanism that enables dynamically adding vCPUs to a VM during runtime, allowing a VM's computation resource to be scaled on demand. When a vCPU hotplug occurs, the host first injects a signal to the VM telling it that a new vCPU is available. The guest OS in the VM will then bring up the vCPU, set up vCPU with schedulers, interrupt handlers, etc., as it would do to a secondary vCPU during boot time. After the new vCPU is brought up, tasks will be balanced to it to utilize the vCPU~\cite{mwaikambo:2004:linux}.
However, in CVMs, the vCPU hotplug mechanism is not supported currently on mainstream CVM platforms due to hardware restrictions or engineering efforts~\cite{tdx-no-hotplug, sev-no-hotplug, cca-no-hotplug}. Intel TDX, AMD SEV, and SEV-ES explicitly states that CPU hotplug is unsupported, requiring all vCPUs within the same CVM to be configured simultaneously~\cite{tdx-no-hotplug,sev-no-hotplug}. ARM CCA has no plans to support hotplug~\cite{cca-no-hotplug}, while AMD SEV-SNP has proposed an extension to support vCPU hotplug, though it remains in the early stages without adequate software support.

\subsection{Serverless Computing}

\bheading{Kubernetes.} Kubernetes (K8s)~\cite{kubernetes:2024:website} is an open-source container platform designed to automate the deployment, scaling, and management of containerized applications. K8s offers a robust suite of tools for container management, including load balancing, automatic resource scheduling, and dynamic scaling. It provides a unified environment to manage diverse workloads efficiently, ensuring applications run reliably and efficiently across a cluster of containers. K8s supports various mainstream container runtimes, including Docker~\cite{docker:2024:website}, Kata~\cite{kata:2024:website}, containerd~\cite{containerd:2024:website}, and CRI-O~\cite{crio:2024:website}.

\bheading{Apache OpenWhisk.} 
Apache OpenWhisk~\cite{apache:2024:openwhisk} is an open-source serverless computing platform designed to execute functions in response to events, dynamically scaling by managing containers based on event intensity. OpenWhisk supports multiple programming languages, including JavaScript and Python. 
There are several key components in OpenWhisk. The \textit{nginx entry point}~\cite{nginx:2023:nginx} serves as the initial gateway, handling SSL termination and forwarding requests. The \textit{controller} acts as the central orchestration component, managing all aspects of HTTP requests, maintaining load balancing, authentication, and directing the execution of actions. The \textit{invoker} triggers functions and containers. Lastly, the foundation of the system is formed by \textit{container clusters} which provide a secure and isolated environment for executing functions.

Deploying OpenWhisk on K8s is a common and recommended approach for managing containers and serverless functions~\cite{apache:2024:openwhisk_on_k8s,baldini:2017:serverless}. This integration leverages K8s' robust resource management and scheduling capabilities. In this setup, K8s manages the infrastructure and container runtime, while OpenWhisk handles incoming events and triggers the execution of corresponding functions, ensuring efficient and scalable serverless computing.

\bheading{Confidential Container (CoCo).} CoCo~\cite{coco:2024:github} allows a trusted execution of containers protected by CVM, such as AMD SEV and Intel TDX. It can be deployed transparently with the upstream K8s. A critical component of CoCo is the use of Kata containers~\cite{kata:2024:website}, which employ CVM to create lightweight, secure VMs for container runtimes. This setup ensures that each containerized application within CoCo is effectively isolated from both the host system and other containers.
Slow cold boot time is one of the major concerns for confidential containers. Previous research suggests that the cold start time for confidential containers may take more than 20 seconds~\cite{segarra:2024:coco}.

%% file: 3-overview.tex
\section{\workcpu Design and Abstraction}
\label{sec:overview}

\subsection{Elastic CVM Overview} 
% \sx{How about title as `Overview'?}
Transforming the current fixed-size CVM into a runtime elastic CVM model can bring numerous potential benefits, making it a win-win situation for both CVM users and CSPs. For CVM users, this change can enable cost-effective usage of CVMs and the ability to handle workload bursts efficiently. For CSPs, it can attract more users and provide hypervisors with the capability to manage CVM resources with finer granularity. Several key research questions about elastic CVM designs are summarized in this section. 

\bheading{How can we efficiently and dynamically adjust the size of CVMs?}
% \begin{center}
% \begin{minipage}[t]{0.99\columnwidth}
% \vspace{-15pt}
% \begin{mdframed}
% \textbf{Q1:} How can we efficiently and dynamically adjust the size of CVMs?
% \end{mdframed}
% \vspace{-10pt}
% \end{minipage}
% \end{center}
For VMs, memory size and the number of vCPUs are the primary resources being utilized. 
In this paper, we mainly focus on dynamically adjusting the number of vCPUs, which can potentially benefit a wide range of applications, especially those that are CPU-intensive. 
% However, current CVM designs do not support runtime vCPU adjustment because the hypervisor is no longer trustworthy.
To enable secure and efficient vCPU runtime adjustment, we introduce a special vCPU type called \textit{\textbf{\workcpu}}.
% to enable elastic CVM size adjustment.
In our design, \workcpus remain in a dormant status, consuming only a few memory pages and no physical CPU resources until they are activated by high workloads. Once the workloads decrease, \workcpus can also return to their dormant status again. This dynamic design allows the CVM size to be adjusted according to workload intensity with minimal resource overhead. Furthermore, \workcpu design can also support \textit{hypervisor-assisted} scheduling. When the CVM and hypervisor agree on the payload to be executed on each \workcpu, the hypervisor can make scheduling decisions for \workcpus almost independently. This approach differs from vCPU hotplug and simple vCPU backup mechanisms and achieves a faster response time.

When it comes to memory, the untrusted hypervisor cannot arbitrarily adjust the CVM's memory size, as this could compromise CVM integrity by maliciously modifying memory mappings, leading to page table manipulation attacks~\cite{morbitzer:2018:severed,morbitzer:2021:severity_new}. 
However, memory size adjustment is theoretically feasible with assistance from the trusted module in Intel TDX or the guest CVM itself in AMD SEV-SNP. Therefore, such adjustments can be combined with the \workcpu design, though the implementation details are beyond the scope of this paper. \looseness=-1

\subsection{\workcpu Design Goals}
Several design goals complement a secure and efficient \workcpu design, including:

\begin{enumerate}[label=\textbf{G\arabic*.},fullwidth,itemindent=0pt,listparindent=\parindent,itemsep=0ex,partopsep=0pt,parsep=0ex]

\item \textbf{Elastic vCPU Number Adjustment and Scalable Performance.}\label{goal_performance} The \workcpu design must facilitate a dynamic adjustment of CVM's CPU capacity, ensuring that performance scales efficiently with actual demand.

\item \textbf{Fast Reaction.}\label{goal_offload} The \workcpu design should enable quick responses to changes in workload intensity.

\item \textbf{Resource Efficiency.}\label{goal_resource} The
\workcpu design must utilize physical resources efficiently, avoiding situations where resources are left idle or pre-allocated without being used.

\item \textbf{Uncompromised CVM Security.}\label{goal_security} The \workcpu design must preserve the existing security level of CVMs, while minimizing code changes to avoid significantly increasing the size of the Trusted Computing Base (TCB).

\item \textbf{Compatibility with Commercial CVM.}\label{goal_compatibility} The \workcpu design must be compatible with commercial CVMs without requiring changes to hardware infrastructures. This ensures a broader adoption of the new elastic model.

\end{enumerate}

\subsection{Threat Model}
In this paper, we considers a similar threat model to most cloud TEEs and CVMs. In this model, the CVM owner deploys CVM instances on a public cloud platform. The cloud service provider (CSP), also known as the hypervisor, is considered untrustworthy or could be compromised, and is interested in accessing secrets within CVM instances. The hypervisor is assumed to have the capability to execute arbitrary privileged instruction sequences that could potentially compromise the confidentiality or the integrity of CVM instances. 
% This includes actions such as pausing, resuming, altering CPU affinity, and launching or deleting hypervisor-controlled CVMs. 
% Despite these threats, \workcpu design must ensure that the protection of both the confidentiality and integrity of the CVMs will not be compromised.
It is important to note that Denial-of-Service (DoS) attacks are beyond the scope of this threat model, and we assume that the hypervisor will not maliciously downgrade the performance of the CVM through any means, such as misconfiguring Non-uniform memory access (NUMA) or CPU affinity. This is because that issues like unstable connections, poor performance, or crashes can be easily detected by the CVM owner, prompting them to discontinue using this CSP. Additionally, ensuring that application implementations running inside the CVM are invulnerable, such as being free from buffer overflow or cache side-channel attacks, is the responsibility of the CVM owner and is not covered by this threat model.

\subsection{\workcpu Abstraction Layer}
\begin{figure}[t]
\centering
\includegraphics[width=0.95\columnwidth]{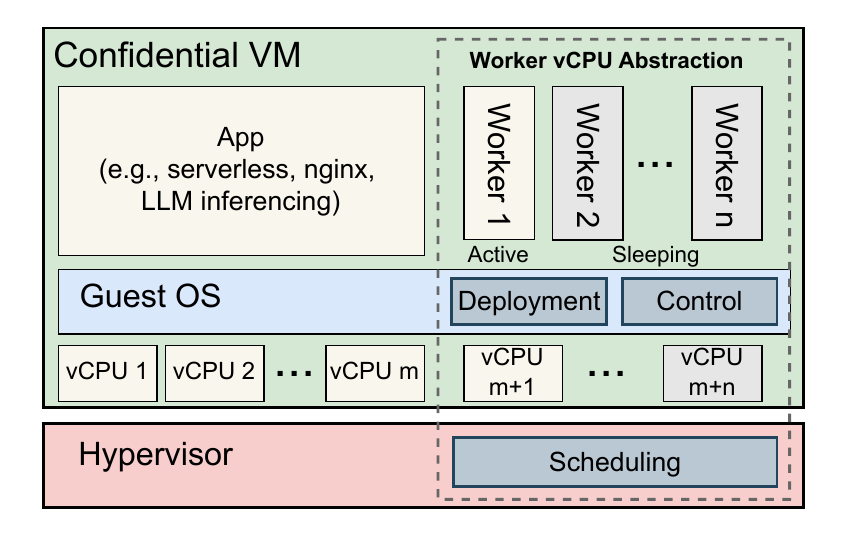}
\vspace{-10pt}
\caption{\workcpu abstraction in CVM architecture.}
\label{fig:overview}
\vspace{-10pt}
\end{figure}

To optimize the use of \workcpus, we introduce an \textit{\workcpu Abstraction Layer} to standardize key components and streamline their integration with existing CVM architecture.
As shown in \figref{fig:overview}, this layer guides \workcpu scheduling, identifies suitable applications, and defines deployment and management processes.

\subsubsection{\textbf{\workcpu Scheduling}} \workcpu transitions between two states: dormant and active. Efficient scheduling strategies may vary based on the target application. The question arises: \textbf{How should \workcpus be scheduled?}

% \begin{center}
% \begin{minipage}[t]{0.99\columnwidth}
% \vspace{-15pt}
% \begin{mdframed}
% \textbf{Q2:} How should \workcpus be scheduled?
% \end{mdframed}
% \vspace{-10pt}
% \end{minipage}
% \end{center}

% \begin{center}
% \vspace{-5pt}
% \minibox[rule=1pt,pad=1pt]{
% \begin{minipage}[h]{0.9\columnwidth}
% {\it \textit{Q2: How should \workcpus be scheduled?}} 
% \end{minipage}
% }
% \vspace{-5pt}
% % \vspace{1pt}
% \end{center}

\ignore{
Generally speaking, there are two \workcpu scheduling approaches. 
The first approach is \textit{self-managed} scheduling, where the CVM owner independently decides whether to wake up the \workcpus or keep them dormant; The second approach is \textit{hypervisor-assisted} scheduling, where the hypervisor monitors the CVM's load and manages the activation of the \workcpus.

Self-managed scheduling can be achieved by implementing a load scheduler inside the CVM to wake up \workcpus, similar to how vCPU hotplugging is done with non-TEE VMs. 
Such a strategy can broadly benefit all types of applications and is particularly beneficial for short-term computational applications, where the CVM is aware of the workload intensity and adjusts the number of parallel threads before each computation.

% However, such strategy usually requires significant changes to the OS and applications. This is because both the OS and the application must be aware of vCPU adjustment and act promptly to fully utilize the newly added vCPUs. The OS must setup the vCPUs and schedule tasks to them; The application must know the new vCPUs are ready and spawn new threads for them correspondingly. Moreover, when the load decreases, the newly added vCPUs need to be removed to save the resource. This also means that both the application and the OS must take care of vCPU removal during runtime as well.

The second strategy is an \textit{auto-regulated} approach, where the scheduling tasks are offloaded to the hypervisor, allowing the hypervisor to make scheduling decisions during runtime. Although this auto-regulated scheduling may limit the range of applicable applications, it is a preferred management method in cloud computing nowadays, enabling faster response in managing \workcpus without the need for manual resource management.
Therefore, in the rest of the paper, we primarily consider systems that use the auto-regulated \workcpu scheduling strategy.
}

Generally speaking, there are two \workcpu scheduling approaches. 
The first approach is \textit{self-managed} scheduling, where the CVM owner independently decides whether to wake up the \workcpus or keep them dormant; The second approach is \textit{hypervisor-assisted} scheduling, where the scheduling is offloaded to the hypervisor to observe the load of the CVM and to launch the \workcpusnospace.

Self-managed scheduling can be achieved by implementing a load scheduler inside the CVM to wake up \workcpus, similar to how vCPU hotplugging is done with non-TEE VMs. 
Such a strategy can broadly benefit all types of applications and is particularly beneficial for short-term computational applications, where the application can easily adjust the number of parallel threads before each computation. However, this strategy typically requires the CVM owner to design a complex scheduler with awareness of the OS and applications, as well as manage vCPU allocation and task scheduling.

% However, such strategy usually requires significant changes to the OS and applications, which can dramatically increase the size of the Trusted Computing Base (TCB). This is because both the OS and the application must be aware of vCPU adjustment and act promptly to fully utilize the newly added vCPUs. The OS must setup the vCPUs and schedule tasks to them; The application must know the new vCPUs are ready and spawn new threads for them correspondingly. Moreover, when the load decreases, the newly added vCPUs need to be removed to save the resource. This also means that both the application and the OS must take care of vCPU removal during runtime as well. 
% \deleted{something} 
% \deleted{Besides, since there are hints of workload-related information by observing \workcpusnospace' activation, it may lead to potential security breaches.}

The second strategy is a \textit{hypervisor-assisted} auto-regulated approach, where the scheduling tasks are offloaded to the hypervisor, allowing the hypervisor to make runtime decisions. The hypervisor observes the overall load of the CVM and adjusts the \workcpus on the fly. While this approach may limit the range of applicable workloads on \workcpus, it remains a preferred method in cloud computing, allowing faster response to manage \workcpus without cloud-user intervention.
Therefore, in the rest of the paper, we primarily discuss the design using the hypervisor-assisted \workcpu scheduling strategy.

% , and can accommodate a significantly large amount of workloads which we will discuss in details in \secref{sec:workloads}.
% Therefore, we designed our system using the hypervisor-assisted \workcpu scheduling strategy.

\subsubsection{\textbf{Applicable Applications with \workcpus}}
\label{sec:workloads}
While the self-managed strategy can broadly benefit all types of applications and the entire CVM, the hypervisor-assisted strategy is less universally applicable, as scheduling is controlled by the untrusted hypervisor. The question arises: \textbf{What types of applications can benefit from the auto-regulated scheduling in \workcpus design?}
% \begin{center}
% \begin{minipage}[t]{0.99\columnwidth}
% \vspace{-15pt}
% \begin{mdframed}
% \textbf{Q3:} What types of applications can benefit from the auto-regulated \workcpus design?
% \end{mdframed}
% \vspace{-10pt}
% \end{minipage}
% \end{center}

There are two key requirements for applicable applications. First, the untrusted hypervisor must be capable of making scheduling decisions, meaning the triggers for \workcpu scaling must be observable by the hypervisor and not hidden by CVM protections. Second, the hypervisor's dynamic adjustment of \workcpus must not impact the functionality of other vCPUs or the overall CVM.

\bheading{Event-Driven Systems and the Producer-Consumer Model.} External event-driven systems, particularly those utilizing the producer-consumer model, are prime examples of applications that can leverage the hypervisor-assisted auto-regulated \workcpus effectively. In this model, producers (incoming workloads to the CVM) are specific external events (e.g., HTTP requests and database queries) that are queued for processing. Consumers (CVM that handles these requests), consisting of a cluster of worker threads, process these requests independently. This setup is inherently adaptable to the auto-regulated model, where each \workcpu can independently serve one worker thread, and the capacity of the entire CVM can be dynamically adjusted by changing the number of active \workcpus.

\bheading{Stateless and Loosely Coupled Worker Threads.} In this model, worker threads are typically stateless or loosely coupled, meaning they have no dependencies on each other and do not retain data between executions. As a result, these threads can be freely put into a dormant state or re-activated based on workload intensity.
A wide range of real-world applications and scenarios fit these application requirements. This includes Function as a Service (FaaS) or serverless (e.g., AWS Lambda, Azure Functions), cloud elastic computing services (e.g., AWS Auto Scaling, Google Cloud Autoscaling), event-driven architecture (e.g., Kafka, RabbitMQ), and common server applications (e.g., Nginx, memcached, and Redis).

\subsubsection{\textbf{\workcpu Deployment}} 

\workcpu deployment mainly consists two phases: initialization during CVM launch and worker thread deployment. The question arises: \textbf{How are \workcpus initialized?}

% \begin{center}
% \begin{minipage}[t]{0.99\columnwidth}
% \vspace{-15pt}
% \begin{mdframed}
% \textbf{Q4:} How are \workcpus initialized?
% \end{mdframed}
% \vspace{-10pt}
% \end{minipage}
% \end{center}

When initializing the CVM, the owner must specify the number of regular vCPUs ($m$) and the maximum number of allowed \workcpus ($n$) to the hypervisor. The hypervisor then launches a CVM with $m + n$ vCPUs.
Specifically, guest kernel modifications are necessary to be aware of \workcpus in terms of scheduling and resource allocation. 
The hypervisor also needs to distinguish \workcpus and regular vCPUs. For example, vCPU $[1, m]$ (regular vCPUs) should always remain in an active status no matter the intensity of the workload. vCPU $[m + 1, m + n]$ (\workcpus) should be in a dormant status after a successful deployment until being waked up by high workloads. 
In the whole procedure, the CPU hardware is not aware of \workcpus and will not distinguish \workcpus from other regular vCPUs. Thus, a normal remote attestation procedure can be used to ensure the identity and the security of the CVM after launching.   

The CVM also needs to configure and bind application threads to each \workcpunospace. The application thread deployment can be done when CVM launches the target application. In general, the target application should initialize a proper number of worker threads based on a vCPU number of $m+n$ instead of $m$. Each \workcpu should be assigned with one worker thread afterwards. 
Modifications or specific configurations in the application scheduler layer may be necessary during thread deployment. 
For instance, certain applications might need to be aware of worker threads running on long-lived regular vCPUs and threads on \workcpusnospace, in order to set CPU affinities. 
Note that all these modifications are not related to the functionality of the target application. For example, in systems that provide serverless function services, all modifications stay in the underlying invoker or configurations, and are not related to the functions executed by cloud users.

\vspace{-10pt}
\subsubsection{\textbf{\workcpu Runtime Control}}
\workcpu control allows for seamless activation and suspension of \workcpus while maintaining the security of both \workcpus and CVMs during runtime. The question arises: \textbf{How can the Worker vCPU be managed during runtime without compromising the security of CVMs?}

% \begin{center}
% \begin{minipage}[t]{0.99\columnwidth}
% \vspace{-15pt}
% \begin{mdframed}
% \textbf{Q5:} How can the \workcpu be managed during runtime without compromising the security of CVMs?
% \end{mdframed}
% \vspace{-10pt}
% \end{minipage}
% \end{center}

Efficient runtime management of \workcpus requires secure coordination between CVMs and the hypervisor. This is crucial because the CVM's internal state is hidden from the untrusted hypervisor, which is responsible for scheduling \workcpus based on the CVM's load. If the hypervisor arbitrarily stops a \workcpu which is in the middle of an execution, the execution may never finish. Therefore, CVMs and \workcpus need to expose certain information to the hypervisor for effective management. For example, an \workcpu could send a ``check-in" signal to the hypervisor to indicate the current tasks are complete, preventing the hypervisor from deactivating it mid-execution. A simple CVM-hypervisor protocol can be predefined for such runtime communication.
To maintain the security guarantee of a CVM, \workcpus are protected at the same level as regular vCPUs. When \workcpus are dormant, all vCPU registers and states are encrypted and stored in the VM Control Block (VMCB), using the identical \texttt{VMEXIT} for regular vCPUs. When \workcpus are active, they operate exactly like regular vCPUs. A detailed discussion of the security implications of \workcpu design and CVM-hypervisor coordination is provided in~\secref{sec:discussion:security}.

%% file: 4-details.tex
\section{\sysname: Serverless Prototype}
\label{sec:serverless}
In this section, we present a serverless platform prototype called \sysname to demonstrate how the \workcpu abstraction layer can help build a system that benefits from \workcpus and the elastic CVM model. The system is named \sysname because it has the ability to fully transform into anything (different serverless functions) it perceives and to turn into a stone (dormant state) when it is inactive for efficiency.

\begin{figure*}[!tbp]
    \centering
    \includegraphics[width=0.985\textwidth]{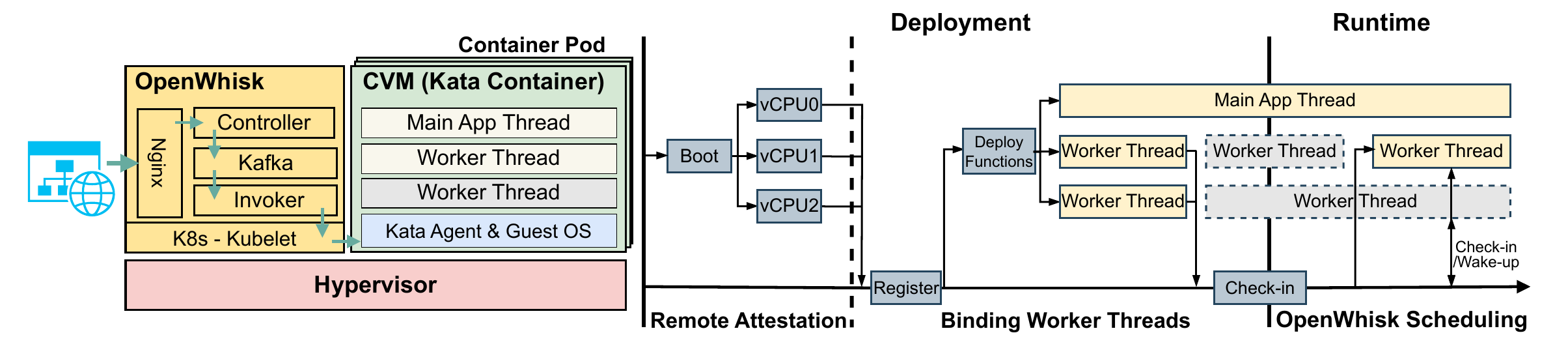}
    \vspace{-10pt}
    \caption{\sysname system components and the workflow.}
    \vspace{-10pt}
    \label{fig:sys}
\end{figure*}

\subsection{\sysname Overview}
\label{sec:impl:overview}
An overview of \sysname is presented in ~\autoref{fig:sys}. \sysname is built on a popular open-source confidential serverless platform configuration, consisting of Apache OpenWhisk~\cite{apache:2024:openwhisk}, Kubernetes (K8s), and Confidential Containers~\cite{coco:2024:github}. Specifically, OpenWhisk orchestrates serverless functions, manages scheduling, and handles triggers and actions. Kubernetes is responsible for the lower container orchestration, scaling, and managing container deployments. Confidential Containers, utilizing CVMs to host containers, provide lightweight, CVM-protected environments for hosting serverless functions.
% , ensuring security against privileged attackers.

Such configurations and systems align with the applicable applications for \workcpusnospace. Specifically, the system is an event-driven architecture utilizing a producer-consumer model for structuring the events queue and comprising some stateless or loosely coupled worker threads. Moreover, the scheduling strategy follows the auto-regulated approach, where the primary scheduling is managed by OpenWhisk running on the untrusted hypervisor. This allows for hypervisor-assisted dynamic adjustment of \workcpus and scalable performance in response to workload intensity.

Compared to a serverless platform that utilizes existing fixed-size CVMs, \sysname offers three main advantages: (1) \textit{Higher Performance.} \sysname addresses workload bursts by waking up \workcpus instead of launching new CVMs. This approach significantly reduces the delay caused by the process of launching clean and brand new CVMs, copying necessary function data, and performing remote attestation. 
(2) \textit{Better Resource Efficiency.} Unlike the fixed-size CVM solution, which may need to pre-allocate CVMs to reduce reaction latency but incurs the cost of maintaining a lot of active vCPUs and reserved VM memory, \sysname occupies minimal physical resources when dormant. Dormant \workcpus require only a few memory pages to store the vCPU register values, making this design highly cost-efficient.
(3) \textit{Finer-grain Resource Management.} \sysname provides hypervisors with the ability to manage CVM resources with finer granularity. For instance, when there is a slight increase in workload, the system can leave the hypervisor to activate \workcpus. If the \workcpus are insufficient to handle the workload, the system can then opt to launch a new CVM.

\subsection{\workcpu Deployment}
\label{sec:impl:deploy}
In \sysname, all containers are hosted by CVMs with the \workcpu feature enabled. Similar to the original configurations in confidential Kata containers, every cloud user has their own K8s pod, which may contain one or more CVMs. Each CVM can host multiple containers, but only containers belonging to the same user can share a CVM or be reused. 
There are three major changes in the deployment process compared to deploying confidential Kata containers with fixed-size CVMs:

% Furthermore, each container hosts only one function at a time.

\bheading{Launching CVMs.} In \sysname, each CVM is configured with $m + n$ vCPUs during initialization, where $m$ vCPUs are regular vCPUs and $n$ vCPUs are \workcpus. The remote attestation procedure is identical to that used in confidential containers, and the CPU hardware recognize each CVM as a normal CVM with $m + n$ vCPUs. Following the remote attestation procedure, the security and integrity of the initial states of all $m + n$ vCPUs, as well as the memory of the CVM, can be guaranteed and verified. A CVM-hypervisor \workcpu protocol (more details in \secref{sec:impl:control}) should also be shared between the hypervisor and CVM owner during this stage.\looseness=-1

\bheading{\workcpu Registration.} After the remote attestation procedure and before the CVM is fully booted, the CVM kernel needs to identify and isolate \workcpusnospace. Specifically, we use a special CPU bitmap mask, called \textit{cpu\_worker\_mask}, to indicate whether a certain vCPU is a \workcpu. 
This special vCPU mask also acts as a ``allow list'', ensuring that only target applications aware of the \workcpus will attempt to schedule worker threads on them after the kernel makes necessary checks. Therefore, when sleeping \workcpus, there is no need to migrate other applications' threads from \workcpus as the only task running on a \workcpu is the application's worker task. This allows the hypervisor to put \workcpus to sleep with no risks of breaking the semantic inside the CVM. Kernel functions that may be scheduled on any CPU, such as the workqueue subsystem, IRQs and housekeeping tasks, are also kept away from \workcpus to ensure that \workcpus can be put into a dormant status without being interrupted or woken up unexpectedly.
After initialization, the CVM registers regular vCPUs and \workcpus in the hypervisor through the \texttt{REG\_MAIN} and \texttt{REG\_WORKER} messages defined in the CVM-hypervisor protocol.

\bheading{Binding Worker Threads.} When launching the target application, $m + n$ worker threads will be created instead of $m$. Among which, $n$ worker threads will be bond to the \workcpu with exactly one thread per \workcpu. %In \sysname, each CVM is configured with $1+3$ vCPUs, consisting of one regular vCPU and three \workcpus.
After the binding is complete, the worker thread should be the only active thread on the \workcpus. Each worker thread then sends a \texttt{CHECKIN} message defined in the CVM-hypervisor \workcpu protocol to inform the hypervisor that it is currently free of tasks and ready to sleep. The hypervisor then puts the \workcpu in a dormant status.

\subsection{\workcpu Scheduling}
\label{sec:impl:schedule}
In \sysname, the scheduling of \workcpus is handled by the untrusted hypervisor. The scheduler relies on observation metrics and scheduling algorithms to make decisions.

\bheading{Observation Metrics.} In CVM environment, the range of observable metrics is more limited compared to settings in non-TEE VMs. This is primarily because CPU states and memory content are now hidden from the untrusted hypervisor, limiting its ability to directly observe these resources. However, the hypervisor is able to have an overall view of the CVM's load by observing the CPU time. Besides, in external event-driven systems, the untrusted hypervisor can still gauge the current workload intensity by observing external events.

% The active vCPUs of a CVM consists of all the regular vCPUs and active \workcpus. 

In \sysname, we choose to observe the CPU load of the CVM as an example. 
The CPU load can serve as a straightforward indicator of the CVM's workload intensity. To calculate the load on the hypervisor side, we use the CPU time information from the \texttt{task\_struct} of each active vCPU to determine how long these vCPUs have been running since the last sampling. We then divide the sum of the active vCPUs' CPU times by the sampling interval to obtain the percentage and use this percentage to represent the current load of the CVM.
Alternatively, the number of encrypted HTTPS requests can serve as a metric. Although the content is encrypted, the hypervisor can estimate workload levels by tracking the number and throughput of these requests. Another useful metric is the approximate processing time for each request. Despite the content of serverless functions being secured by the CVM, the hypervisor can infer processing times by monitoring the interval between when a CVM receives a request from a specific IP address and when it returns the processed result.

%

% \todo{Add how load is calculated here}

%is designed to report its status to the hypervisor through the \texttt{CHECKIN} message in the CVM-hypervisor \workcpu protocol, indicating if they are idle to prevent the hypervisor from incorrectly putting an active \workcpu to sleep.

%indicating if the \workcpu is not performing any tasks at this point and is ready to sleep% to prevent the hypervisor from incorrectly putting an active \workcpu to sleep.

\bheading{Scheduling Algorithms.} In \sysname, we implement a simple yet dynamic strategy for scaling by monitoring the CPU load of the CVM. If the total load of the active vCPUs reaches a certain pre-defined threshold, the hypervisor then wakes up a \workcpu. If the load of active vCPUs drops below a certain pre-defined threshold, the hypervisor can then request a \workcpu to sleep upon the next \texttt{CHECKIN} message in the CVM-hypervisor \workcpu protocol.  This \texttt{CHECKIN} message is issued by a \workcpu to report that it has finished the previous serverless function and has not yet pick up the next function, which is a checkpoint that the hypervisor can put the \workcpu to sleep without disrupting CVM operations. If all \workcpus are occupied and the workload continues to exceed capacity, additional CVMs are deployed to scale up further through OpenWhisk's invoker.

In practical deployments, CSPs typically use more sophisticated algorithms to enhance system performance. Although comprehensive scheduling algorithms are beyond the scope of this paper, different algorithms should generally be adoptable in \sysname, as they operate entirely on the hypervisor side.

\subsection{Runtime Control}
\label{sec:impl:control}
Runtime control ensures that the dynamic scheduling of \workcpus does not compromise the functionality or security of the target application and CVM. In addition, it provides necessary control and assistance for efficient scheduling.

\begin{table}[!t]
    \centering
    % \setlength{\abovecaptionskip}{0mm} 
    % \setlength{\belowcaptionskip}{0mm}
    % \hspace*{12pt}
    \begin{adjustbox}{scale=0.9}
    \small
    \centering
    \begin{tabularx}{\columnwidth}{l|>{\raggedright\arraybackslash}X}
    \hline
    \hline
    \textbf{Message} & \textbf{Description} \\       
    \hline
    \texttt{REG\_MAIN} & Register this vCPU as a regualr vCPU. \\ \hline
    \texttt{REG\_WORKER} &  Register this vCPU as a \workcpu.   \\ \hline
    \texttt{CHECKIN} & Notify the hypervisor that this \workcpu is idle and can sleep. \\ \hline
    \texttt{DEREG\_WORKER} & Deregister this \workcpu from the hypervisor, e.g., when shutdown the CVM.  \\ \hline
    \texttt{DEREGISTER} & Deregister this entire VM from the hypervisor when shutting down.  \\ \hline
    \hline
    % \bottomrule 
    \end{tabularx}
    \end{adjustbox}
    \caption{A simple CVM-Hypervisor protocol example.}
    \label{tab:protocol}
    \vspace{-10pt}
\end{table}
\bheading{\workcpu Status Transitions.} 
In \sysname, the untrusted hypervisor manages the scheduling of \workcpu, making the necessary isolation of \workcpu essential to avoid crashes that could be caused by sleeping a \workcpu. \sysname uses the special \textit{cpu\_worker\_mask} to distinguish \workcpu and generally follows the requirements of the CPU hotplug~\cite{mwaikambo:2004:linux} to modify the guest kernel. Unlike CPU hotplug, which requires proactively moving application threads from a vCPU that is about to sleep, \workcpu provides a more isolated vCPU environment, restricting threads that can run on \workcpu. 
% In this design, \workcpu is not scheduled to run any application threads other than the target application's worker threads. 
This minimizes code changes related to the complex scheduler and allows for rapid \workcpu status transitions.
Similar to CPU hotplug requirements~\cite{mwaikambo:2004:linux}, additional modifications include I/O interrupts redirection, pausing internal local timer ticks, and modified Inter-Processor Interrupts (IPI). For instance, when waking up a \workcpu from a dormant status, \workcpu itself enforces a local TLB flush to prevent local and stale Translation Lookaside Buffers (TLBs).

Furthermore, the design of the executor inside the CVM strictly adheres to the producer-consumer pattern, featuring stateless and loosely coupled worker threads. This means there are no locks or dependencies between different worker threads upon issuing the \texttt{CHECKIN} message. As a result, the hypervisor's control over the starting and stopping of worker threads does not impact other busy worker threads and will not compromise the overall functionality of the CVM.

\bheading{CVM-hypervisor \workcpu Protocol.} 
A simple CVM-hypervisor protocol is pre-defined for the necessary communication between the CVM and the hypervisor. This communication is facilitated by certain pre-defined \texttt{CPUIDs}. In AMD SEV-ES, \texttt{CPUID} instructions are trapped by the VC handler~\cite{amd:2020:ghcb} inside the guest kernel. The guest kernel performs necessary checks, copies the information to a shared guest-host communication block (GHCB), and then invokes the hypervisor. \tabref{tab:protocol} lists an example of a basic CVM-hypervisor protocol. The messages \texttt{REG\_MAIN} and \texttt{REG\_WORKER} are used during VM startup. The messages \texttt{DEREG\_WORKER} and \texttt{DEREGISTER} are mainly used during VM shutdown. To achieve a graceful shutdown, the CVM will need to deregister all of its \workcpus and then deregister the VM from the management of the hypervisor-assisted scheduling. The hypervisor can also actively teardown a CVM as usual in the case of a malicious VM or VM crashes.
The \texttt{CHECKIN} message is used at runtime to inform the hypervisor whether the current worker vCPU is idle.

It is worth noting that the CVM-hypervisor protocol can also be extended to enable additional \workcpus control from the CVM side. For example, the CVM can use this protocol at runtime to inform the hypervisor whether to start or stop a specific \workcpunospace, or to adjust the maximum number of \workcpus allowed to be active.

%% file: 5-evaluation.tex
\section{Evaluation}
\label{sec:evaluation}
In this section, we conduct a micro evaluation of \workcpu design and a macro evaluation of the \sysname prototype. 

\subsection{Evaluation Setup}
Our testbed is built atop an AMD workstation equipped with an AMD EPYC 7251 8-Core Processor, 64 GiB of RAM, a 1 TiB disk, and supports AMD SEV-ES protection. We use Ubuntu 22.04 and implemented our system using Linux Kernel 6.6.18. On the guest kernel side, we only added $\sim$400 Line of Code (LoC) which makes the TCB increment minimal. On the hypervisor kernel side, we added $\sim$600 LoC. The QEMU and OVMF are all straight out-of-the-box meaning that the kernel modifications are the only changes in the infrastructure level.
Since our servers do not support the SNP feature, each CVM has been configured with SEV-ES protection for demonstration purposes. However, the design of the \workcpu is not constrained by the version of the CVM and can be seamlessly ported to an SEV-SNP environment. 
%The guest CVM kernel was also forked and compiled from the same AMD GitHub repository. 

%The necessary software for running an SNP-protected VM was sourced directly from AMD SEV's official GitHub repository~\cite{amd:2023:github} (specifically from the sev-snp-devel branch\footnote{Commit: \texttt{fbd1d07628f8a2f0e29e9a1d09b1ac6fdcf69475}}). This includes the host kernel (from the sev-snp-iommu-avic\_5.19-rc6\_v4 branch), QEMU (from the snp-v3 branch), and OVMF (from the master branch).

To evaluate our system, we present results from a synthetic benchmark and a real-world integration with the OpenWhisk serverless platform. The synthetic benchmark presented in \secref{sec:evaluation:synthetic} is tested in a controlled setting to show performance and overheads of the \workcpu design. The OpenWhisk integration presented in \secref{sec:evaluation:openwhisk} demonstrates how our system benefits real-world applications.

%For the serverless prototype, related configurations are as follows: \todo{Add configurations related to the serverless}

\subsection{Synthetic Benchmarks}
\label{sec:evaluation:synthetic}

% \ignore{
% \begin{figure}
%     \centering
%     \includegraphics[width=.45\textwidth]{figures/exec_timeline.pdf}
%     \caption{Timeline of a synthetic workload on our system versus a regular VM.}
%     \label{fig:exec-timeline}
% \end{figure}
% }

% \begin{figure}
%     \centering
%     \includegraphics[width=.5\textwidth]{figures/exec_timeline2_v2.pdf}
%     \caption{Timeline of a synthetic workload on our system versus a regular VM.}
%     \label{fig:exec-timeline}
% \end{figure}

\begin{figure}[t]
\centering
\begin{subfigure}[b]{0.495\columnwidth}
\includegraphics[width=\textwidth]{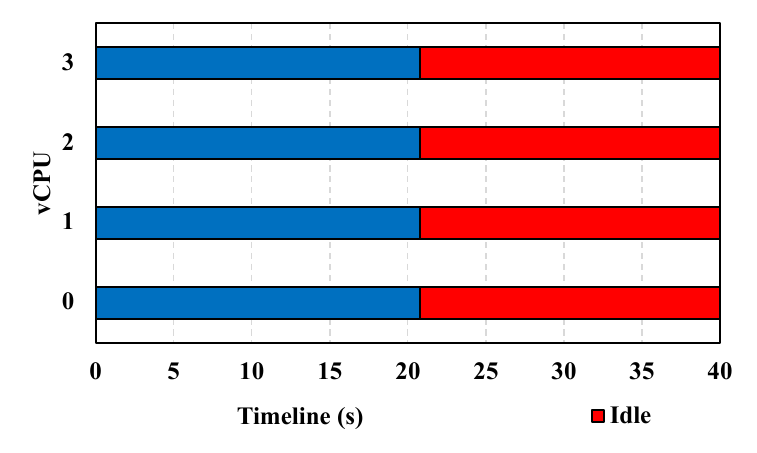}
\caption{4-vCPU Regular VM.}
\label{fig:micro:0s}
\end{subfigure}
\begin{subfigure}[b]{0.49\columnwidth}
\includegraphics[width=\textwidth]{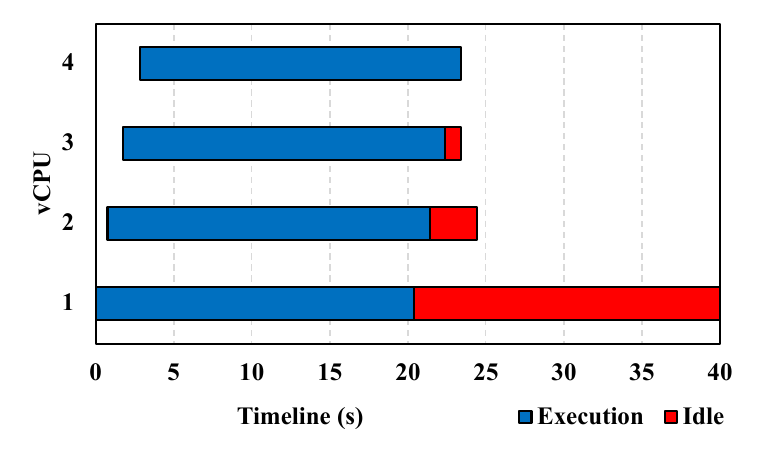}
\caption{\sysname(sampling rate 0.5s).}
\label{fig:micro:05s}
\end{subfigure}
\begin{subfigure}[b]{0.49\columnwidth}
\includegraphics[width=\textwidth]{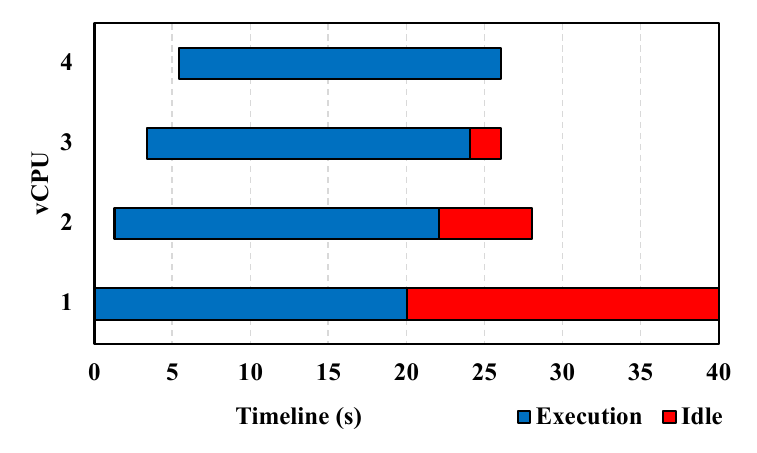}
\caption{\sysname(sampling rate 1s).}
\label{fig:micro:1s}
\end{subfigure}
\begin{subfigure}[b]{0.49\columnwidth}
\includegraphics[width=\textwidth]{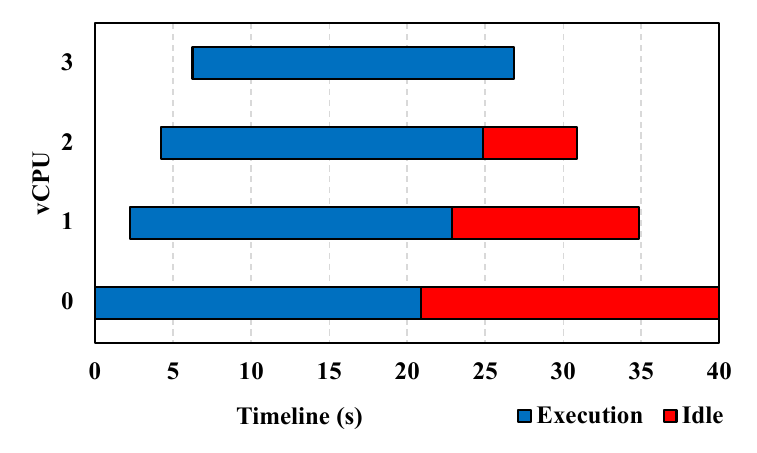}
\caption{\sysname(sampling rate 2s).}
\label{fig:micro:2s}
\end{subfigure}
\caption{Timeline of a synthetic workload on \sysname versus a regular VM. }
\label{fig:exec-timeline}
\vspace{-10pt}
\end{figure}

We first provide insights of our system using a synthetic workload. We used a 4-vCPU CVM for evaluation. Among the four vCPUs, we chose three vCPUs as \workcpus. The workload involves calculating the Fibonacci sequence for 20 seconds, repeated four times. This means a sequential execution should take around 80 seconds, but it can also be completed in four parallel 20-second executions.
Additionally, we explored the impact of the sampling rate of the scheduler in the hypervisor. We tested our system against a straight-forward parallel execution and illustrate the result in \autoref{fig:exec-timeline}.

\bheading{Performance and CPU Resource Benefits.} As shown in the execution results in \autoref{fig:micro:0s}, a regular CVM with 4 active vCPUs completes the task in exactly 20 seconds and then stays idle.
In our system, performance depends on the sampling rate, which refers to the time interval at which the scheduler makes the decision to wake up or put a \workcpu to sleep. For example, with a sampling rate of 2 seconds, the scheduler in the hypervisor adjusts the number of active \workcpus every 2 seconds.
As shown in \autoref{fig:exec-timeline}, the entire execution takes 27 seconds with a sampling interval of 2 seconds, 25 seconds with a sampling interval of 1 second, and 22 seconds with a sampling interval of 0.5 seconds. We can see that the delay generally decreases as the sampling frequency increases. While the latency of waking up a \workcpu can be negligible, an extremely high sampling rate may not benefit performance. This is because the hypervisor needs to frequently sample the current workload, which can introduce significant overhead and potentially degrade overall system efficiency. Consequently, finding an optimal sampling rate is crucial and should depend on the application. 
% lead to a slight increase in the delayed cycle, as the \workcpu may take a small amount of time to recover after returning from the hypervisor. \added{unclear,latency = 0}

% While this scheduling strategy introduced slightly delay in our system, it is still 3x faster than the sequential execution that takes 80 seconds.

\bheading{Epilogue Latency.} One of the key benefit of our system is that idle \workcpus can be taken away by the hypervisor and saving CPU resources. From the idling result in \autoref{fig:exec-timeline}, we can see that after execution, the \workcpus gradually returned to dormant instead of still being active but in a idle way. Since the regular vCPU's load ended earlier, the sampler in hypervisor noticed this before the load on vCPU 3 finished and therefore vCPU 3 was put to sleep immediately. Other \workcpus were also put to dormant one by one aligning to our scheduling strategy. We can also see that the higher the sampling rate, the faster \workcpus gets back to dormant after the load. Compared with a regular VM which has all four vCPUs active after the execution, we have only one regular vCPU active after four sampling cycles.
% \added{Add a proportion number of idle time,maybe in the figure?}

\bheading{System Overheads.} In our system, there are two potential overheads. One is the VM boot-time overhead, the other one is the sampler's overhead in the hypervisor. For the VM's boot overhead, a regular VM takes 9.05 seconds to boot while our system takes 10.36 seconds. This 1 second overhead was likely to be imposed by disabling interrupts and workqueues on the \workcpus since these works would now need to be handled by the regular vCPU. For the sampler's overhead, it takes 20 $\mu$s to sample each VM. When the sampling interval is set to 2 seconds, the cost is around 0.01\%, which is negligible. Even when the sampling interval is set to 0.5 second, the cost is still only 0.04\%.

\bheading{vCPU Add/Removal Latency.} vCPU add/removal latency refers to the time it takes for a vCPU to begin responding or enter sleep mode after a scheduling decision is made. We compared the latency in our system with that of vCPU hotplugging in non-CVM environments to highlight our efficiency and low latency.
In our design, the isolation of work vCPUs allows the hypervisor to quickly put them to sleep or wake them up for tasks. The primary latency is due to \texttt{VMENTER} and \texttt{VMEXIT} operations. In contrast, vCPU hotplugging/hotunplugging in non-TEE VM environments involves complex coordination between the VM OS and hypervisor, including communication via interrupts, configuring the vCPU and moving threads.

\begin{figure}
    \centering
    \includegraphics[width=.45\textwidth]{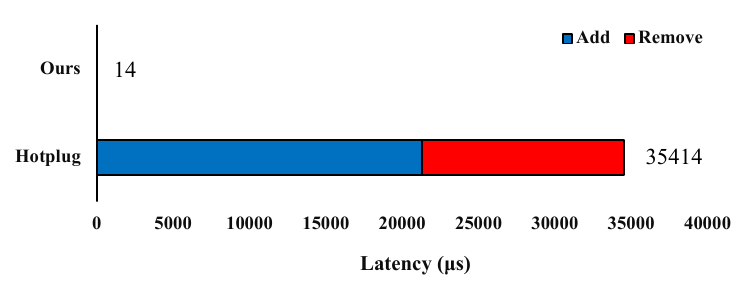}
    \vspace*{-10pt}
    \caption{\workcpu add/removal latency versus conventional vCPU hotplug/hotunplug.}
    \label{fig:hotplug-latency}
    \vspace{-15pt}
\end{figure}

We tested both and measured the latency of each method in \autoref{fig:hotplug-latency}. As one can see, the average latency of our design for the round-trip add and removal is around 14 $\upmu$s while the conventional vCPU hotplug/hotunplug can cost 35,414 $\upmu$s, which is 2,530x slower than ours. This is because the guest OS will have to setup the vCPU when hotplug and migrate all the threads when hotunplug.
Note that for vCPU hotplug/hotunplug method, the latency does not include overheads for application awareness. In practical settings, the application needs to be aware of this and setup additional threads, the cost of it can be even higher. In contrast, our system requires no runtime awareness of the application, and it only takes 14 $\upmu$s to begin responding even in a real-world setting.

\subsection{Application Evaluation with \sysname}
\label{sec:evaluation:openwhisk}

To highlight the real-world benefits of our design, we implement and evaluate our \sysname prototype, demonstrating its superior performance and efficiency compared to a vanilla OpenWhisk system.
In serverless computing, the cold start problem significantly increases latency, reduces throughput, and adds overhead~\cite{serverless-fact3}. This issue is exacerbated in a confidential setting, where the boot cost of a TEE can be substantial~\cite{reusable-enclaves}. This means that when a user's function suddenly experiences a peak invocation, the system will have to boot multiple confidential containers to handle the demand since the CVM hosting the container cannot scale, even if the security policy permits execution within the same CVM, such as when the invocations originate from the same user.

\bheading{Experimental Setup.} In \sysname, the CVM hosting the container can have multiple \workcpus available, allowing it to scale dynamically based on the load. We modified OpenWhisk by adding additional worker threads within the confidential container for each CVM and binding them to \workcpus. We also assume and configure a security policy that permits OpenWhisk to allocate workloads to the same container VM when the functions are invoked by the same user. Such configurations are common in commercial cloud serverless platforms~\cite{wang:2018:peeking}. The baseline for comparison is a standard confidential serverless setup, configured as a vanilla version of OpenWhisk running on confidential Kata containers (CoCo).

\begin{figure}[t]
    \centering
    \includegraphics[width=.42\textwidth]{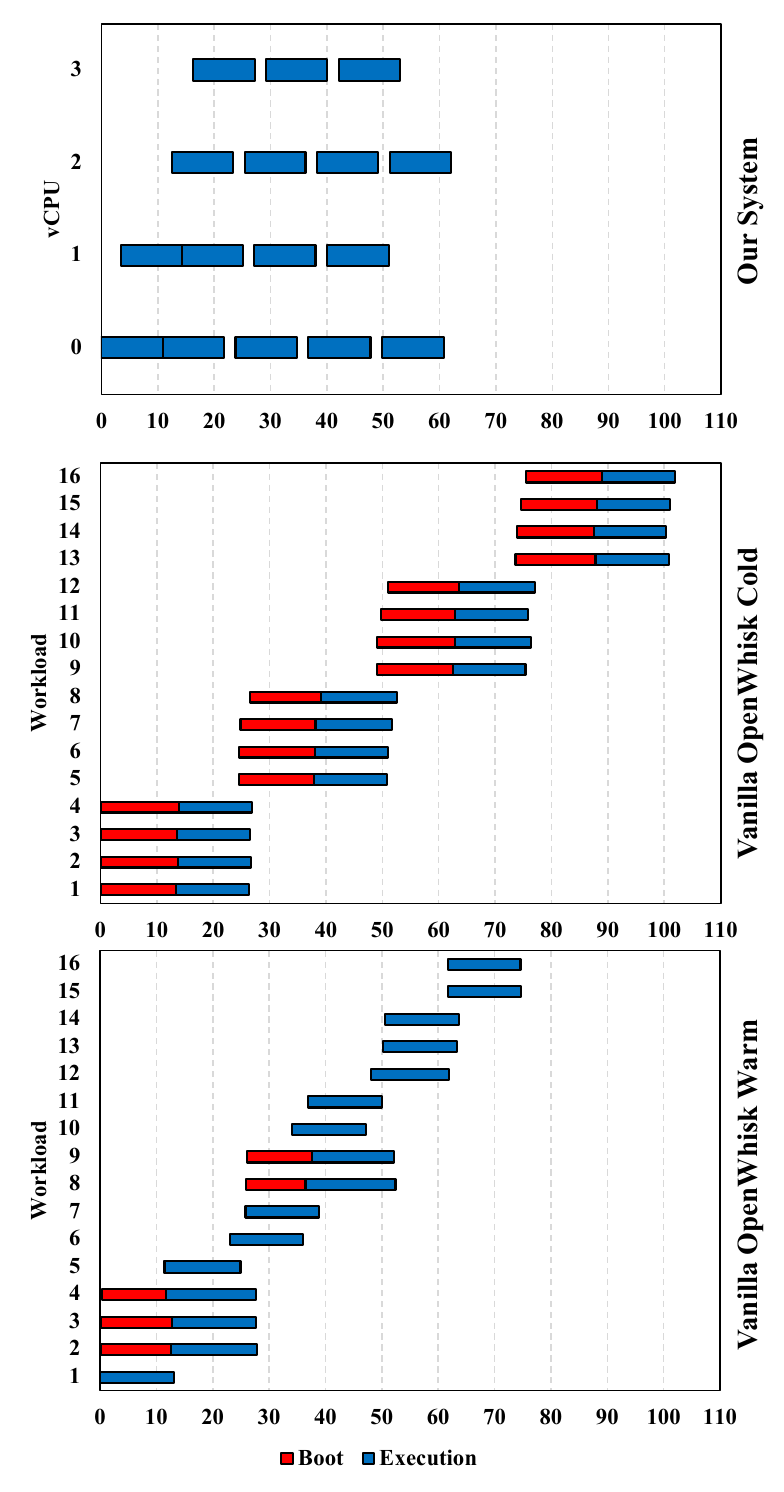}
    \vspace*{-10pt}
    \caption{Timeline of workloads executing on OpenWhisk using \sysname and a vanilla system (with/without warm-keeping). }
    \vspace{-10pt}
    \label{fig:openwhisk-eval}
\end{figure}

We applied a constant load of approximately 10 seconds using the prime number algorithm from~\cite{openwhisk-bench}. To simulate burst demand, we issued 16 concurrent loads to the system. Both systems are configured with identical maximum CPU capacity. In our system, we deployed one container with 4 vCPUs, consisting of one regular vCPU and three \workcpus. The scheduling sampling rate was set to 2 seconds. For the vanilla version, we configured it to support a maximum of 4 containers (1 vCPU each container). The default and hard-coded setting of the kata container is one container per CVM. Thus, a maximum of four CVMs may be generated during runtime in the vanilla version. 
To show the implication of warm start versus cold start, we also tested the vanilla version of OpenWhisk with both settings by enabling (as default) or disabling the warm-keeping feature. By keeping containers warm, we are simulating that the peak workloads are all invocations to the \textit{same} function (\eg, containers can reused). Otherwise, the peak workloads are invocations of \textit{different} functions. The results are illustrated in \figref{fig:openwhisk-eval}. 

\bheading{Performance Comparison.} We can observe that both our system and the vanilla OpenWhisk utilized all 4 available cores. For our system, \workcpus scale accommodate the loads. The hypervisor samples the load of the system then dynamically and gradually increase the \workcpus to the system. Thus, all 16 workloads finishes in 61.995 seconds.

However, for the cold start vanilla OpenWhisk, since the boot time of a confidential container can take around 15 seconds and existing containers cannot scale, the vanilla OpenWhisk system will need to create new containers and swallow the overheads. In this scenario, each container must be booted from scratch, causing massive delays. The overall time consumption for executing the 16 workloads is 101.883 seconds, 64\% slower than our system. 
In the warm-keeping vanilla OpenWhisk setting, although it is faster than the cold start version, it still suffers from the cold start problem when a high load suddenly comes in, as demonstrated in workloads 2-4. This is due to the limited number of pre-warmed containers. Additionally, the OpenWhisk scheduler sometimes opts to start new containers, as seen in workloads 8 and 9, which introduces additional overhead. The total time for the warm-keeping vanilla OpenWhisk to complete the 16 workloads is 74.628 seconds, making it 20\% slower than our system.\looseness=-1
% \added{efficiency is not defined here}\sx{See now}

\begin{figure}
    \centering
    \includegraphics[width=.45\textwidth]{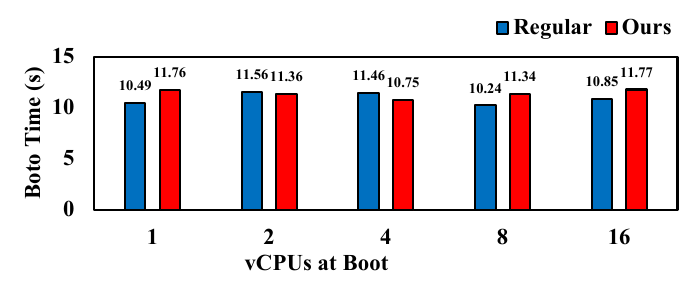}
    \vspace{-10pt}
    \caption{OpenWhisk CVM Boot-time in \sysname and the vanilla system with different numbers of vCPUs.}
    \label{fig:openwhisk-boot}
    \vspace{-15pt}
\end{figure}

\bheading{Efficiency Comparison.} In addition to performance benefits, our system demonstrates better \textit{CPU Efficiency} compared to the vanilla OpenWhisk. We define CPU efficiency as the ratio of busy CPU time (execution time) to the total time, including cold boot overhead, reflecting how efficiently the system uses CPU resources.
In our system, the \workcpus enter a dormant state when idle and handle tasks immediately upon waking, ensuring that all CPU time of the vCPUs is fully dedicated to execution ($\scriptstyle\sim$100\%) after the deployment. In contrast, the vanilla OpenWhisk must allocate CPU time for booting processes, which reduces its overall efficiency.
For the cold start scenario, the efficiency for the 16 workloads was only 49.4\%, due to the fact that no container is ready to use until a cold start. However, even in the warm-keeping scenario, the efficiency improved to 78.9\%, but it was still significantly lower than our system.
% For our system, since the \workcpus are put to sleep on idle and handles the work immediately on wake up, the CPU time of the vCPUs is all spent on the execution. However, for the vanilla OpenWhisk have to spend the CPU time on booting, the efficiency degrades when compared with our system. For the cold start scenario, the efficiency of the 16 workloads was 49.4\% which the cold start time was even higher than the execution itself. For the warm-keeping scenario, the efficiency was 78.9\%, while higher than the cold start setting, was still significantly lower than our system.

\bheading{Boot-Time Overheads.} To assess the additional setup required during the CVM's booting procedure in our system, we tested the boot-time overheads for different vCPU configurations and compared them with a vanilla executor container of OpenWhisk using CoCo. The results are presented in~\figref{fig:openwhisk-boot}. The data shows that, regardless of the number of vCPUs or whether the system is our modified kernel version or the vanilla version, the boot-time overheads were consistently around 11 seconds. This indicates that the number of vCPUs does not significantly impact the overall boot time. Furthermore, it demonstrates that our system does not introduce any noticeable additional boot overheads.

\bheading{Scheduling and Memory Size Strategy.} Determining the most efficient scheduling strategy and observation metrics (\eg, CPU load or I/O rate) for scheduling \workcpus is beyond the primary scope of our paper. In real-world cloud platforms, CSPs can leverage existing serverless scheduling strategies to further refine this process. Thus, our prototype evaluation primarily focuses on demonstrating that \workcpus can be quickly started and stopped in a serverless platform, and how this capability optimizes the efficiency of CPU resources. Similarly, the selection of an appropriate CVM memory size, which is highly application-dependent, is discussed in~\appref{sec:appendix:mem_size}.

%% file: 6-discussion.tex
\section{Discussion}
\label{sec:discussion}

\subsection{Security Analysis}
\label{sec:discussion:security}

In this section, we analyze the security of the \workcpu design throughout its entire lifecycle, including the security implications of a malicious hypervisor.

\bheading{Remote Attestation.} The design of \workcpu does not require the CPU hardware to be aware of or differentiate between \workcpu and regular vCPUs. Consequently, CVMs with \workcpu enabled can utilize the same remote attestation procedure as regular CVMs to ensure the security of the CVM's initial state. This includes the verifiability of the entire CVM's initial memory and states of all vCPUs (including both \workcpus and regular vCPUs). 
All internal behaviors of the CVM post-attestation are then protected by the runtime protection enforced by CVM TEE design. 

% Similarily, Within the \workcpu design, after remote attestation, CVMs with \workcpu enabled register the \workcpu and run target applications. These activities are also safeguarded by the standard protections of a normal CVM.

\bheading{Confidentiality of CVMs with \workcpus.} 
The confidentiality of \workcpus during runtime is safeguarded similarly to that of a regular vCPU within a CVM. When active, the CPU states of the \workcpu are protected just as those of any regular vCPU. For example, on AMD SEV-ES platform, during each \texttt{VMEXIT}, the register values of the \workcpu are encrypted, stored, and integrity-protected within the Virtual Machine Save Area (VMSA). Furthermore, all memory accessed by the \workcpu is protected through the memory encryption techniques.
For transitions in \workcpu status and the CVM-hypervisor protocol, the \workcpu proactively sends a \texttt{CHECKIN} message to the hypervisor when idle, signaling readiness to transition to a dormant state. All register values, with the exception of the 4 GPRs that transmit pre-defined protocol messages via the \texttt{CPUID} instruction, are encrypted and securely stored in the VMSA by the trusted CPU hardware through the atomic \texttt{VMEXIT} instruction. Similarly, upon resuming from a dormant state, the CPU hardware retrieves all register values from the VMSA via an atomic \texttt{VMRUN} instruction, ensuring complete confidentiality of all secret register values.

\bheading{Additional Information Leakage.} The \workcpu design might introduce some information leakage through interactions with the hypervisor. However, we argue that this exposure is minimal, as the hypervisor could already infer most of this data through other means. The hypervisor could primarily gain access to two types of information during interactions. 

Firstly, the hypervisor might be able to guess the type of program running on \workcpus. This is because the CVM owner may choose to inform the hypervisor of the triggering conditions when deploying \workcpus. These conditions could include high CPU utilization, such as over 90\%, or specific I/O events, like receiving network packets from a designated IP address. Such details might inadvertently reveal certain natures of tasks executed within the CVM. For example, in our prototype serverless platform, the hypervisor could conclude that each \workcpu is handling a distinct serverless function. 
Despite of this, we argue that similar inferences could be made even without \workcpus, as a malicious hypervisor can exploit established side channels — such as those based on controlled interrupts~\cite{li:2021:cipherleaks,wilke:2023:sev}, page tables~\cite{morbitzer:2018:severed,morbitzer:2021:severity_new}, or performance counters~\cite{werner:2019:severest} — to fingerprint application activity. Importantly, the security of the data processed by the applications remains protected.
Secondly, the hypervisor can assess whether the \workcpu is idle by the \texttt{CHECKIN} message. This message provides the hypervisor with indirect insights into the CVM’s operational load. we contend that such leakage does not introduce an incremental risk in CVM. Even without \workcpus, the hypervisor can already assess a CVM’s workload using other observable metrics, such as CPU utilization of each vCPU or I/O event frequency.

\bheading{Integrity of CVMs with \workcpus.} 
When \workcpus are in an active status, the integrity of tasks running on \workcpus are protected by CVM's TEE design.
During status transitions of \workcpus, the integrity of the CVM is maintained as all worker threads on the \workcpus are either stateless or loosely coupled. 
Malicious actions by the hypervisor, such as improperly starting, stopping, or delaying a \workcpu, will not compromise the integrity of worker threads on other vCPUs. Such actions may, however, result in a DoS attack if the \workcpu is not idle. In such cases, the CVM owner can detect the resulting poor performance and can choose to discontinue services with the CSP.
A more detailed discussion on the potential impacts of existing attacks on CVMs and their implications is provided in~\secref{sec:related:security}.

\bheading{TCB Size Increment.} The guest kernel support for \sysname should remain within a minimal range. To support basic functionality of \workcpu design, we only added $\sim$400 LoC to the guest kernel, making the TCB size increase negligible. Other modifications to OpenWhisk and KVM do not count for TCB increment as they are outside of the CVM. \looseness=-1

\bheading{Malicious or Vulnerable \workcpu.} We also discuss the security risks of a malicious \workcpu. Since tasks on \workcpus are not verifiable by the hypervisor, malicious CVM owners could manipulate task states, such as falsely triggering or reporting sleep readiness, neither of which benefits the CVM and could lead to operational faults. Despite these risks, \workcpus remain isolated from the hypervisor through hardware virtualization (e.g., KVM).

% the potential security implications of a malicious \workcpu on the hypervisor. Tasks running on \workcpus are not verifiable by the hypervisor, creating a risk of manipulation by malicious CVM owners who might provide incorrect information. For instance, they could inappropriately trigger a \workcpu or falsely report it as ready to sleep, neither of which benefits the CVM and could lead to operational faults. Despite these risks, \workcpus are isolated from the hypervisor through hardware-enabled virtualization technologies, ensuring no direct security threats to the hypervisor.

% Similarly, it is the responsibility of the CVM owner to ensure that programs deployed on \workcpus strictly adhere to the \sysname's design for safely running stateless workloads and are free from bugs such as buffer overflow.

% \bheading{Fixed memory mapping and size.} The size of CVM memory and its mapping (mapping information between VM physical address and system memory physical address) are typically fixed. Adjusting the memory size or performing a memory swap during runtime is time-consuming and necessitates validation by the hardware. For instance, a requires the CVM to re-validate this information using the \texttt{PVALIDATE} instruction in AMD SEV~\cite{amd:2020:snp}. 

\subsection{vCPU Scaling Methods Comparison} 
% \sx{Can we make this in one line?}
Apart from the \workcpu design, two alternative methods can also achieve CVM scaling at runtime: (1) \textit{vCPU hotplugging} (which is not yet supported in CVMs, as discussed in \secref{sec:background:hotplug}), and (2) \textit{pre-deployed backup vCPUs}. However, both methods suffer from trade-offs shown in \autoref{tab:comparison}. 

Hotplug vCPU can achieve dynamic scaling and allow the removal of vCPUs when unnecessary, thus it consumes no CPU resources at idle and maintains idling efficiency. However, this approach requires both the guest OS and the application to be aware of vCPU add/removal events to respond accordingly. As a result, the hypervisor cannot freely adjust or schedule vCPUs unless it utilizes a similar CVM-hypervisor protocol, as introduced in this paper. Moreover, hotplug vCPUs suffer from slow response time as discussed in~\secref{sec:evaluation:synthetic}.  
% In addition, the hot plug/hotunplug procedure itself is very heavy~\cite{vcpu-hotplug-overhead}

Simply pre-deploying some backup vCPUs and application threads when initializing can also help handle burst workloads during runtime. However, without the \workcpu abstraction proposed in this paper, all scheduling can only be done by the guest OS. This approach results in low efficiency because backup vCPUs cannot dynamically scale up or down and may consume resources even under no load, such as managing timer interrupts and performing Linux housekeeping tasks.

In contrast, the \workcpu design dedicates each \workcpu to a task and ensures separation between normal vCPUs and \workcpus. They don't consume CPU time when idling, allowing the hypervisor to make scheduling decisions with minimal dependency, resulting in highly efficient performance with minimal response time.

%% file: 7-related.tex
\section{Related Work}
\label{sec:related}
In this section, we discuss existing work about CPU adjustment and the implications of known attacks against CVM.

\subsection{CPU Ballooning}
CPU ballooning~\cite{uhlig:2004:towards} is a technique employed in virtualized environments to dynamically adjust the vCPU allocation based on demand. 
This approach, similar to CPU hotplug, faces challenges such as accommodating worker queues and scheduling. In our \workcpu design, we address these by using pre-allocated vCPUs designated for specific applications, significantly reducing the latency associated with vCPU state transitions within a CVM environment to negligible.

Previous research mainly has leveraged vCPU ballooning to address the \textit{Double Scheduling} issue.
Song~\etal~\cite{song:2013:schedule} introduced a system named VCPU-Bal. VCPU-Bal assigns an identical total number of vCPUs and physical CPUs (pCPUs). This one-to-one mapping of vCPUs to dedicated pCPUs can thus eliminates the \textit{Double Scheduling} problem.
VCPU-Bal allows the dynamic adjustment of vCPU number per VM, controlled by a global scheduler. This scheduler determines the number of vCPUs by dividing each VM's weight with the total weight from all VMs. When deactivating a vCPU using CPU hotplug feature, a vCPU Ballon Module inside VM will clean the runqueue, move alive threads to other vCPUs, and deregisters any associated events in this vCPU.
VCPU-Bal's performance is assessed using some computational tasks, such as dedup, streamcluster, and canneal from PARSEC~\cite{bienia:2008:parsec}, which vary the number of worker threads in accordance with the number of vCPUs. Results indicate that for those tasks, VCPU-Bal can outperform virtualized platforms that over-commit vCPUs beyond the number of available pCPUs. 
Miao~\etal,~\cite{miao:2015:flexcore} later extend VCPU-Bal and propose FlexCore. FlexCore further reduces the vCPU hotplug latency and explores the time breakdown in how VCPU-Bal mechanism can reduce the execution time. 

% \begin{table}[!t]
%     \setlength{\abovecaptionskip}{10pt}   %标题与下文距离
%     \setlength{\belowcaptionskip}{0pt}   %标题与图距离
%     % \setlength\tabcolsep{3pt}
%     % \hspace*{12pt}
%     \begin{adjustbox}{scale=0.99}
%     % \footnotesize
%     \small
%     \centering

%     % \resizebox{1.0\linewidth}{!}{
%     \begin{tabularx}{\columnwidth}{l|>{\centering\arraybackslash}X|>
%     {\centering\arraybackslash}X|>
%     {\centering\arraybackslash}X} 
%     % \toprule
%     \hline
%     \hline
%     \textbf{Properties}   & \textit{Hotplug vCPU } & \textit{Backup vCPU} & \textit{Worker vCPU} \\       
%     % \midrule
%     \hline
%     Dynamic Scaling  & \Circle & \xmarkn & \Circle \\ \hline
%     No Runtime Awareness Req'd & \xmarkn & \LEFTcircle & \Circle \\ \hline
%     Scheduling Offloading  & \LEFTcircle & \xmarkn & \Circle \\ \hline
%     %No Double Scheduling  & \xmarkn & \xmarkn & \Circle \\ \hline
%     Fast Response on Demand  & \xmarkn & \LEFTcircle & \Circle \\ \hline
%     Minimal CPU Time Waste & \Circle & \xmarkn & \Circle\\ 
%     % Cost Efficiency & \Circle & \xmarkn & \Circle\\ 
%     \hline
%     \hline
%     % \bottomrule 
%     \end{tabularx}
%     \end{adjustbox}
%     \caption{Comparison with other conventional vCPU scaling methods. \Circle~means yes. \LEFTcircle~means partially. \xmarkn~means no.} %\ZQ{Since you have Y*, why don't you use circle, fullcircle, and halfcircle?}}\sx{Revised} \mengyuan{if you check 7.1, double scheduling has a special meaning. Does it fit here, may or may not, I am not sure}
%     \label{tab:comparison}
%     \vspace{-10pt}
% \end{table}

\begin{table}[!t]
    \setlength{\abovecaptionskip}{10pt}   %标题与下文距离
    \setlength{\belowcaptionskip}{0pt}   %标题与图距离
    % \setlength\tabcolsep{3pt}
    % \hspace*{12pt}
    \begin{adjustbox}{scale=0.99}
    % \footnotesize
    \small
    \centering

    % \resizebox{1.0\linewidth}{!}{
    \begin{tabularx}{\columnwidth}{l|>{\centering\arraybackslash}X|>
    {\centering\arraybackslash}X|>
    {\centering\arraybackslash}X} 
    % \toprule
    \hline
    \hline
    \textbf{Properties}   & \textit{Hotplug vCPU } & \textit{Backup vCPU} & \textit{Worker vCPU} \\       
    % \midrule
    \hline
    Dynamic Scaling  & \Circle & \xmarkn & \Circle \\ \hline
    No Runtime Awareness Req'd & \xmarkn & \Circle & \Circle \\ \hline
    Scheduling Offloading  & \LEFTcircle & \xmarkn & \Circle \\ \hline
    %No Double Scheduling  & \xmarkn & \xmarkn & \Circle \\ \hline
    Fast Response  & \xmarkn & \Circle & \Circle  \\ \hline
    % Minimal CPU Time Waste & \Circle & \xmarkn & \Circle\\ 
    Idling Efficiency & \Circle & \xmarkn & \Circle\\ 
    \hline
    \hline
    % \bottomrule 
    \end{tabularx}
    \end{adjustbox}
    \caption{Comparison with other conventional vCPU scaling methods. \Circle~means yes. \LEFTcircle~means partially. \xmarkn~means no.}
    \label{tab:comparison}
    \vspace{-10pt}
\end{table}

Additionally, a recent research paper published in July 2024, titled CPC~\cite{chen:2024:cpc}, explores the application of CPU ballooning to perform hypervisor maintenance tasks, such as live migration, within a CVM environment. 
While CPC focuses on deploying a hypervisor-controlled, Virtual Machine Privilege Level (VMPL)-protected security module inside the CVM for maintenance tasks, this approach introduces a semantic gap between the module and the applications running within the CVM. For instance, the module cannot directly interact with or comprehend the specific context of these applications. In contrast, our work introduces a structured \workcpu abstraction that enhances the execution capability of the CVM's native programs and supports hypervisor-assisted scheduling offloading. More discussion about adding VMPL to \workcpu can be found in \appref{sec:appendix:vmpl}.

% \subsection{TEE CPU Performance Optimization}
% Zhou~\etal~\cite{zhou:2023:core} enhance both the performance and security of CVMs by isolating physical CPU cores. This isolation strategy, known as "core slicing", assigns specific CPU cores, memory, and virtual I/O devices to each guest CVM, thereby mitigating risks associated with data leaks and side-channel attacks while boosting performance. A dedicated slice manager, operating on a reserved CPU core, oversees the allocation and management of these resources. However, this approach departs from the traditional VM-level abstraction provided by conventional hypervisors and consequently restricts the number of CVMs that each physical server can support.

\subsection{Known Attacks Against CVM}
\label{sec:related:security}
Since the introduction of commercial CVM, there has been extensive research into their security. Numerous studies have identified a variety of vulnerabilities in commercial cloud VM designs. A detailed classification of known attacks on CVMs, including side-channel attacks, and their impact on \workcpu design is discussed in \appref{sec:appendix:attacks}.
Among these attacks, the majority have been mitigated through iterative updates. The most relevant attacks concerning \workcpu design are the two newly identified interrupt injection attacks~\cite{schluter:2024:heckler,schluter:2024:wesee}. In these attacks, an adversary can compromise the integrity or confidentiality of vCPUs by maliciously injecting harmful interrupts. To counter these threats, AMD has introduced a ``restricted injection'' solution~\cite{amd:2020:snp}. Specifically, each \workcpu can be configured to use an ``alternate injection'' mode. This mode ensures that virtual interrupt queues are managed directly by the CVM itself, thus safeguarding \workcpus by controlling the injection of interrupts in a manner akin to the protection afforded to regular vCPUs.

%% file: 8-conclusion.tex
\section{Conclusion}
\label{sec:conclusion}
This paper proposes the ``Elastic CVM'' concept and the \workcpu to overcome the limitations of fixed capacities in current CVM designs. By enabling dynamic CPU resource management, our approach improves operational efficiency and resource adaptability without sacrificing security. The \sysname platform demonstrates faster response latency and better resource management in the confidential serverless platform, marking a representative example toward more flexible and cost-effective confidential computing environments.

%% file: paper.bbl

%% file: 0-appendix.tex
\appendix

\section{Memory Sizing Strategy}
\label{sec:appendix:mem_size}
While the paper primarily focuses on CPU resources, the memory capacity of CVMs could also pose a limitation on their performance. However, the appropriate memory size largely depends on the application being run. Thus, determining a suitable memory size does not fall within the scope of this paper. 
For the key applications discussed herein, such as databases, network servers, or ML inference tasks, substantial memory is occupied by storing shared elements like ML models or database content. Typically, the suitable memory size can thus be estimated during the VM deployment stage and does not generally become a bottleneck even during surges in workload.
Moreover, existing CVM designs enable dynamic adjustment of memory size. For instance, in AMD SEV-SNP systems, the guest CVM can utilize the \texttt{PVALIDATE} instruction to effectively reclaim and adjust its memory size during runtime, as described in~\cite{amd:2020:snpapi}.

\section{Adding Internal Isolation using VMPL}
\label{sec:appendix:vmpl}
In the scenarios and the \sysname design considered in this paper, the programs run by the \workcpu, such as a serverless function or a worker thread, are determined by the CVM owner only. Therefore, we design the \workcpus to operate at the same security level as the other vCPUs and did not implement any hardware-enforced isolation between the \workcpus and the rest of the CVM. Many existing research projects use this feature to enable internal isolation within a CVM ~\cite{qiang:2018:se,alder:2019:sfaas,goltzsche:2019:acctee}.
However, we want to point out that if special requirements arise, the system can easily adopt SEV's Virtual Machine Privilege Level (VMPL) feature to further provide internal hardware-enforced isolation. This isolation might be useful when the programs running on the \workcpus are from a third party and could potentially pose a potential threat to the entire CVM. By setting the \workcpus to a lower privilege VMPL level and restricting the memory pages they can access, the \workcpu can be isolated in a small portion of the CVM without the risk of tampering normal vCPUs or the rest of the CVM. 

\section{Classification of Known Attacks on CVMs}
\label{sec:appendix:attacks}
Known vulnerabilities against CVMs range from \textit{unencrypted register values}~\cite{hetzelt:2017:security,kaplan:2017:seves}, \textit{weak or vulnerable memory encryption}~\cite{li:2021:cipherleaks,li:2022:systematic,du:2017:sevUnsecure,wilke:2020:sevurity}, \textit{malicious page mapping manipulation}~\cite{li:2020:crossline,hetzelt:2017:security, morbitzer:2018:severed,morbitzer:2019:extract_new, morbitzer:2021:severity_new}, \textit{vulnerable remote attestation procedure}~\cite{buhren:2019:insecure}, \textit{physical voltage fault injection attacks}~\cite{buhren:2021:one}, \textit{vulnerable interrupt and exception handling}~\cite{schluter:2024:heckler,schluter:2024:wesee}, \textit{vulnerable micro-architecture design}~\cite{li:2021:tlb,Zhang:2024:CacheWarp},  and different \textit{side-channel attacks}~\cite{li:2022:systematic,wilke:2023:sev,wang:2023:pwrleak,werner:2019:severest}.

\bheading{Side channels and controlled channels.} While side-channel and controlled channel attacks are not commonly included in the threat models for many commercial CVMs, such as AMD SEV, Intel TDX, and ARM CCA, evidence from existing side-channel attacks demonstrates their effectiveness in extracting data from applications protected by CVMs to varying degrees. Significantly, CVMs are vulnerable to both traditional side-channel attacks~\cite{werner:2019:severest,li:2022:systematic,wang:2023:pwrleak,werner:2019:severest} and TEE-specific controlled channel attacks~\cite{wilke:2020:sevurity,morbitzer:2021:severity_new,wilke:2021:undeserved,li:2021:tlb,wilke:2023:sev,li:2021:cipherleaks,van:2017:sgx}.
% In terms of threats from side-channel attacks, a malicious hypervisor can exploit side-channel information more effectively and has access to gathering information from privileged instructions or interfaces. 
% For instance, the attacker might use performance counters (PMC)~\cite{werner:2019:severest,li:2022:systematic}, power usage interfaces~\cite{wang:2023:pwrleak}, and special CPU features like instruction-based sampling~\cite{werner:2019:severest} to gather a diverse set of information, thereby increasing the risk of information leakage from within the CVM.
% In terms of controlled channel attacks, 
% a malicious hypervisor can manipulate the execution of a CVM through tactics such as page table manipulation and interrupt injection. For example, 
% an attacker can unset the ``present'' bit in the page table to intercept page accesses~\cite{wilke:2020:sevurity,morbitzer:2021:severity_new,wilke:2021:undeserved,li:2021:tlb}, or use a timer interrupt to force the CVM to exit after a brief interval, allowing for the single-stepping of instruction execution within the CVM~\cite{wilke:2023:sev,li:2021:cipherleaks,van:2017:sgx}.
To mitigate these threats, developers may need to modify the source code of applications running within CVMs to enhance their resistance to these types of attacks. Importantly, the introduction of \workcpu does not inherently worsen the vulnerability to side-channel attacks.

\bheading{Malicious interrupt injection.} Recent studies have identified two attacks where a malicious hypervisor injects harmful interrupts into a CVM, compromising its execution. Schlüter~\etal~\cite{schluter:2024:heckler,schluter:2024:wesee} demonstrate that certain interrupts or exceptions can corrupt the VM if the CVM does not rigorously verify the register values returned by the hypervisor. By injecting these faulty interrupts during the operation of applications within the CVM, an attacker can manipulate the application's control flow or expose sensitive register values. To counteract these vulnerabilities, AMD has introduced a vCPU feature called ``restricted injection''~\cite{amd:2020:snp}, specifically designed to prevent such malicious activities.

\bheading{Unencrypted register values.} In AMD SEV, the register values that are stored within the VM Control Block (VMCB) are not encrypted during \texttt{VMEXIT}. This vulnerability exposes the register values to potential inspection and manipulation by a malicious hypervisor~\cite{hetzelt:2017:security,werner:2019:severest}
Such vulnerability has been fixed since AMD SEV-ES. In SEV-ES and SEV-SNP, register values are encrypted before being stored in the VM Save Area (VMSA) and have additional integrity protection, effectively reducing the risk of such attacks.

\bheading{Vulnerable and weak memory encryption.} Vulnerable memory encryption~\cite{du:2017:sevUnsecure,wilke:2020:sevurity} stems from the potential to reverse engineer the tweak function within the hardware-accelerated memory encryption engine. 
This vulnerability allows an attacker to manipulate the ciphertext stored in memory by replacing or replaying it. 
% However, this issue has been addressed with the introduction of an updated memory encryption engine in the EPYC Zen 2 architecture.
On the other hand, weak memory encryption involves the use of deterministic encryption modes, such as AES-XEX (AMD SEV) or AES-XTS (Intel TDX). These modes are susceptible to attackers that can continuously monitor the ciphertext data stored in the memory. 
Specifically, the attacker can observe the recurrence of identical ciphertext blocks, leveraging this ``ciphertext side channel'' to potentially extract secret keys from sophisticated cryptographic algorithms.

\bheading{Malicious page mapping manipulation.} In AMD SEV and SEV-ES, a malicious hypervisor has the ability to manipulate nested page tables, enabling the redirection of guest physical addresses to alternative memory pages. This can result in the virtual machine accessing incorrect data or instruction pages~\cite{li:2020:crossline,hetzelt:2017:security, morbitzer:2018:severed,morbitzer:2019:extract_new, morbitzer:2021:severity_new}. Such vulnerability is fixed by the newly added reverse map table (RMP table) in SEV-SNP, which securely backs up the correct mapping information between system physical addresses and guest physical addresses and prevents a malicious hypervisor from compromising the integrity of programs running within the CVM by modifying the nested page table. 

\bheading{Physical voltage fault injection.} Physical voltage fault injection attacks~\cite{buhren:2021:one} use physical device to inject voltage fault to AMD SEV. Specifically, during the booting phase of a CPU, an attacker might induce hardware malfunctions to circumvent specific security protocols, thereby allowing the AMD secure processor to run compromised firmware. Such vulnerabilities are generally addressable through firmware updates and more secure protocol design.